\documentclass[journal=ancac3,manuscript=article]{achemso}

\usepackage{chemformula} 
\usepackage[T1]{fontenc} 
\usepackage[version=4]{mhchem}
\usepackage{enumitem}

\NewDocumentCommand{\MakeTitle}{ +m +m +m }{%
    \begingroup
        \renewcommand\thefootnote{\@fnsymbol\c@footnote}%
        \def\@makefnmark{\rlap{\@textsuperscript{\normalfont\@thefnmark}}}%
        \long\def\@makefntext##1{\parindent 1em\noindent
            \hb@xt@1.8em{%
                \hss\@textsuperscript{\normalfont\@thefnmark}%
            }##1%
        }%
        \if@twocolumn
            \ifnum \col@number=\@ne
                \MakeTitleInner{#1}{#2}{#3}
            \else
                \twocolumn[\MakeTitleInner{#1}{#2}{#3}]%
            \fi
        \else
            \newpage
            \MakeTitleInner{#1}{#2}{#3}
        \fi
        \thispagestyle{plain}
    \endgroup
    \setcounter{footnote}{0}%
}

\author{Emilio Méndez}
\author{Rocio Semino}
\email{rocio.semino@sorbonne-universite.fr}
\affiliation[]
{Sorbonne Université, CNRS, Physico-chimie des Electrolytes et Nanosystèmes
Interfaciaux, PHENIX, F-75005 Paris, France.}

\title{Thermodynamic Insights into the Self-assembly of Zeolitic Imidazolate Frameworks from Computer Simulations}


\begin{document}


\begin{tocentry}

\includegraphics[width=8.2cm]{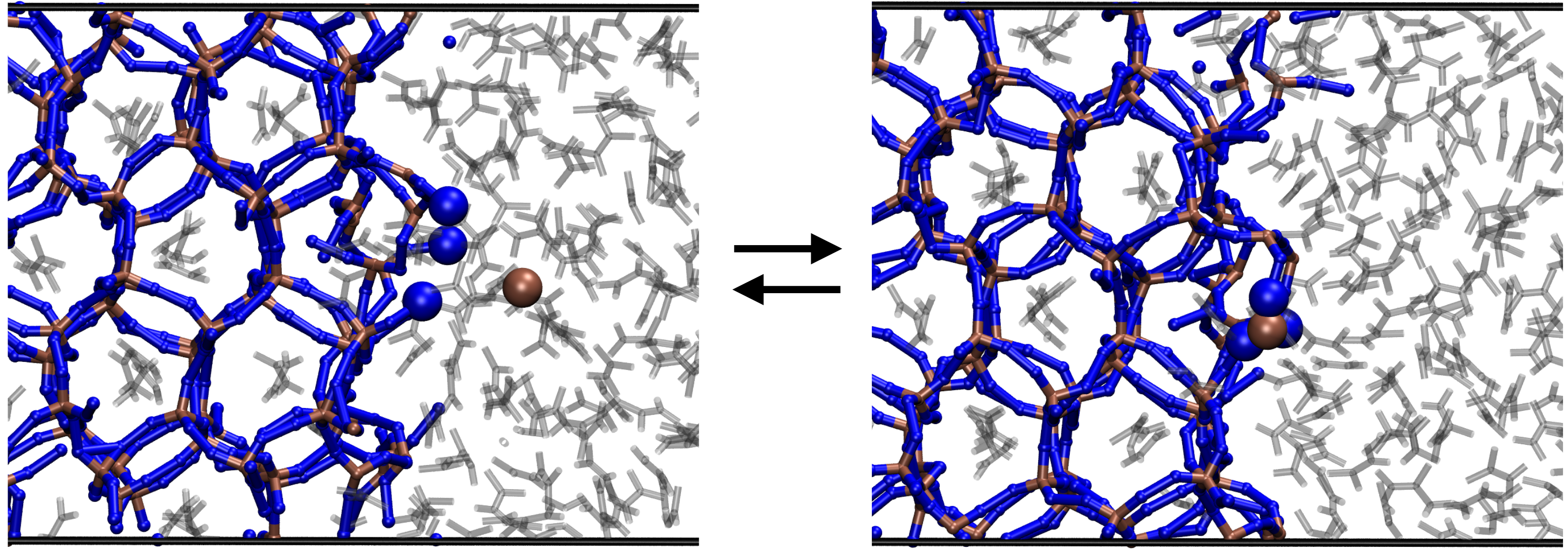}

The molecular level mechanisms and thermodynamics of early nucleation and growth of ZIF-4 are revealed through computer simulations.

\end{tocentry}

\begin{abstract}
New metal-organic frameworks (MOFs) are periodically synthesized all over the world due to the wide range of societally and environmentally relevant applications they possess. However, the mechanisms and thermodynamics associated to MOF self-assembly are poorly understood because of the difficulties in studying such a multi-scale process with molecular-level resolution. In this work, we performed well-tempered metadynamics simulations of the early nucleation and late growth steps of the self-assembly of ZIF-4 using a reactive force field. We found that the formation of building blocks is a complex, multi-step process that involves changes in the coordination of the metal ion. Saturating the ligand coordination of a metal ion is more energetically favorable during growth than during early nucleation.
The addition of a fourth ligand is less exergonic than it is for the first three and the associated free energy is highly dependent on the local environment of the undercoordinated metal ion. The stability of this bond depends on the strength of the solvent--metal ion interaction. Incorporating a ligand to a ZIF-1 crystal is less favorable compared to the more stable ZIF-4 polymorph. Milder differences were found when comparing the growth of (100), (010) and (001) ZIF-4 surfaces.
\end{abstract}

\section{Introduction}

Zeolitic Imidazolate frameworks (ZIFs) are a family of metal-organic frameworks (MOFs) characterized by their high chemical and thermal stability.\cite{Chen2014} These porous solids are formed by a metal ion (typically Zn$^{2+}$, Co$^{2+}$, Fe$^{2+}$ or Cd$^{2+}$) tetrahedrally coordinated to a bidentate imidazolate(Im)-based ligand. The metal ion Zn--N(Im) coordination bonds lead to metal-ligand-metal angles of 145$^\circ$, equivalent to the T-O-T angles encountered in zeolites. As a consequence, ZIFs tend to share topology with some zeolites (\textit{sod} and \textit{rho} are the most common ones). Moreover, the flexibility afforded by these coordination bonds together with the high stability in water that results from their hydrophobicity,\cite{Ortiz2014} confer ZIFs an enormous potential for stimuli-responsive applications.\cite{Iacomi2021} 

Even though almost twenty years have passed since the initial discovery of six Zn(Im)$_2$ porous polymorphs (ZIF-1, ZIF-2, ZIF-3, ZIF-4, ZIF-6 and ZIF-10) by Park and coworkers,\cite{Park2006} their phase diagram, as well as the chemistry underlying their synthesis mechanisms are still largely unknown. These six polymorphs were all synthesized in solvothermal conditions, all with dimethylformamide (DMF) as a solvent, except for ZIF-2, which was made in a 2:1 DMF/N-Methyl-2-pyrrolidone mixture. Reactants were Zn(NO$_3$)$_2$ and imidazole in all cases. The main difference between the synthesis conditions leading to these different polymorphs lies in the metal:ligand and reactants:solvent ratio as well as in the temperature and synthesis times.\cite{Park2006} The size and morphology of a particular polymorph can also be modified by changing synthesis conditions.\cite{Yu2020,Kaneti2017} Clearly, as for many other families of materials, the synthesis process for ZIFs depends on a delicate balance between thermodynamic and kinetic aspects.\cite{Lewis2009,Cheetham2018} In cases where the synthesis is under thermodynamic control, the denser, enthalpically favored phases are typically obtained (including ZIF-zni and ZIF-coi),\cite{Zhang2019,Schrder2013} while when under kinetics control, less stable polymorphs could be favored.\cite{VanSanten1984,Cardew2023} It is worth mentioning though that the enthalpy difference between porous polymorphs can sometimes be very low.\cite{Lewis2009}

The thermodynamically most favorable porous Zn(Im)$_2$ polymorph, ZIF-4, crystallizes in the \textit{cag} topology, which has not yet been found in zeolites. In an in-depth study of the phase diagram of this MOF,\cite{Widmer2019} none of the other five polymorphs found by Park and coworkers were detected. Whether this is a matter of lack of stability of certain polymorphs\cite{BousselduBourg2014} or due to limitations in the experimental measures remains yet to be elucidated. Instead, five other crystalline polymorphs were found, including the less porous, high pressure phases ZIF-cp-II and ZIF-cp-III, the high temperature phases ZIF-hpt-I and ZIF-hpt-II and the dense ZIF-zni form. In addition, two amorphous phases and a liquid-like phase were detected.\cite{Fonseca2021} Yet another crystalline porous phase, ZIF-4-cp was also experimentally obtained in another study,\cite{Henke2018} and its place in the phase diagram has been recently determined via computer simulations.\cite{Mendez2024_2}

Despite the vast amount of experimental and computational efforts to better understand the complex landscape of Zn(Im)$_2$ and other zeolitic MOFs structures,\cite{Eddaoudi2015,Henke2018,VanVleet2018,Widmer2019,Guillerm2019,Freund2021,Wang2022,Barsukova2023,CastilloBlas2024,Mu2024,Lee2023,Mendez2024,Mendez2024_2,Du2024,Shi2024,Castel2024} we still lack fundamental knowledge on the thermodynamics and dynamics of the Zn--N(Im) coordination bonds, which are at the heart of the synthesis and phase transitions of these materials. Understanding what (and why) influences the formation of Zn--N(Im) bonds will bring us closer to the holy grial of ZIF rational design, that has been elusive for decades.

Recent advances in computational techniques have made it possible to model the early stages of self-assembly of MOFs. \cite{Yoneya2015,Biswal2016,Biswal2017,Colon2019,Kollias2019,Wells2019,Filez2021,Balestra2022,Balestra2023}
 Yoneya and coworkers pioneered the field by modelling the complexation between Ru$^{2+}$ and Pd$^{2+}$ and 4,4'-bipyridine within an implicit solvent. Biswal and Kusalik went even farther by adding an explicit solvent and proposing the use of cationic dummy atom models (CDA) to take into account the anisotropy in charge distribution around Zn$^{2+}$ ions to better understand the formation of rings within MOF-2.\cite{Biswal2016,Biswal2017} CDA models were also applied to studying hydration energies in pristine and defective UiO-66,\cite{Su2021} as well as to study the mechanical properties of several MOFs.\cite{Jawahery2019}
Colon and coworkers were the first to apply enhanced sampling methods to accelerate the sampling of the dynamics of coordination bonds thus enabling to reach later stages of the nucleation process.\cite{Colon2019} Kollias and coworkers relied on well-tempered metadynamics\cite{Barducci2008} to explore changes in free energies associated to early nucleation stages of MIL-101(Cr) as a function of the solvent and ionic force.\cite{Kollias2019} Balestra and Semino combined CDA models with well-tempered metadynamics simulations for the first time to study the early stages of the nucleation of ZIF-8.\cite{Balestra2022} Later on, Filez and collaborators have combined simulations and experiments to gain insight into the nucleation mechanism of ZIF-67(Co), focusing on changes in the coordination symmetry of the Co$^{2+}$ ion.\cite{Filez2021} Following a different philosophy based on a Monte Carlo approach, Wells and colleagues have modeled the formation of different polymorphs as a function of the composition of the system.\cite{Wells2019}    

In this contribution, we build on previous works to answer crucial questions related to the thermodynamics and dynamics of coordination bonds. We first focus on the mechanism of formation of a [Zn(Im)$_4$]$^{2-}$ complex in solution, starting from a fully solvated Zn$^{2+}$ ion. We tackle the question on whether the successive ligand additions are equally favorable in terms of the free energy or not, and how does this process compare with adding a new Zn$^{2+}$ or Im$^{-}$ ion onto a preformed ZIF surface, as a model of the growth stages of the self-assembly process. We study the mechanism and thermodynamics of adding these extra-ions to the surface and look at whether they depend on the  ${h k l}$ indices and/or on the polymorph studied. 

\section{Results}

\subsection{Formation of early [Zn(Im)$_n$]$^{2-n}$ complexes}

We start out analysis by studying the fundamental reaction steps that take place at the very beginning of the nucleation process. In particular, we focused on the formation of the [Zn(Im)$_4$]$^{-2}$ complex from the dissociated ions, that will serve as building unit for the further growth of the ZIF structures.
In order to gain insight into the free energy associated to this transformation, we 
ran well-tempered metadynamics simulations in which the ligands reversibly attach and detach to the Zn center thanks to the partially reactive force field nb-ZIF-FF.\cite{Balestra2022} This allowed us to compute free energy differences between all the possible intermediate species that are formed 
along this process. 
These simulations also offer molecular-level information concerning the mechanism of these reactions.

We divided this process into two steps:
        
(I)      \ce{Zn$^{2+}$ + 2 Im$^{-}$ <=> [Zn(Im)$_2$]}

(II)     \ce{[Zn(Im)$_2$] + 2 Im$^{-}$ <=> [Zn(Im)$_4$]$^{2-}$}

This was done to guarantee reasonable convergence times, as a new collective variable is needed to treat each coordination bond, and the cost of a metadynamics simulation scales rapidly with the number of collective variables (CVs) (see Methods section).
The distances between each reactive imidazolate and the Zn ion and the coordination number between the Zn ion and O(DMF) atoms were used as CVs for both steps. The imidazolates that were interchanged in the first step were kept connected to the Zn ion via a harmonic restraint for the addition of the two last ligands. All simulations were performed at the experimental ZIF-4 solvothermal synthesis temperature of 400 K\cite{Park2006}.

\subsubsection{From an octahedral to a tetrahedral Zn ion}

It is known that the most stable Zn$^{2+}$  complex in DMF solution 
comprises six solvent molecules\cite{Ishiguro1990}. On the other hand, the coordination of Zn with imidazolate in all ZIF crystals is tetrahedral. 
In what follows, we delve into the question of how this change of coordination takes place.

\begin{figure}[h]
\centering
\includegraphics[width=1.0\textwidth]{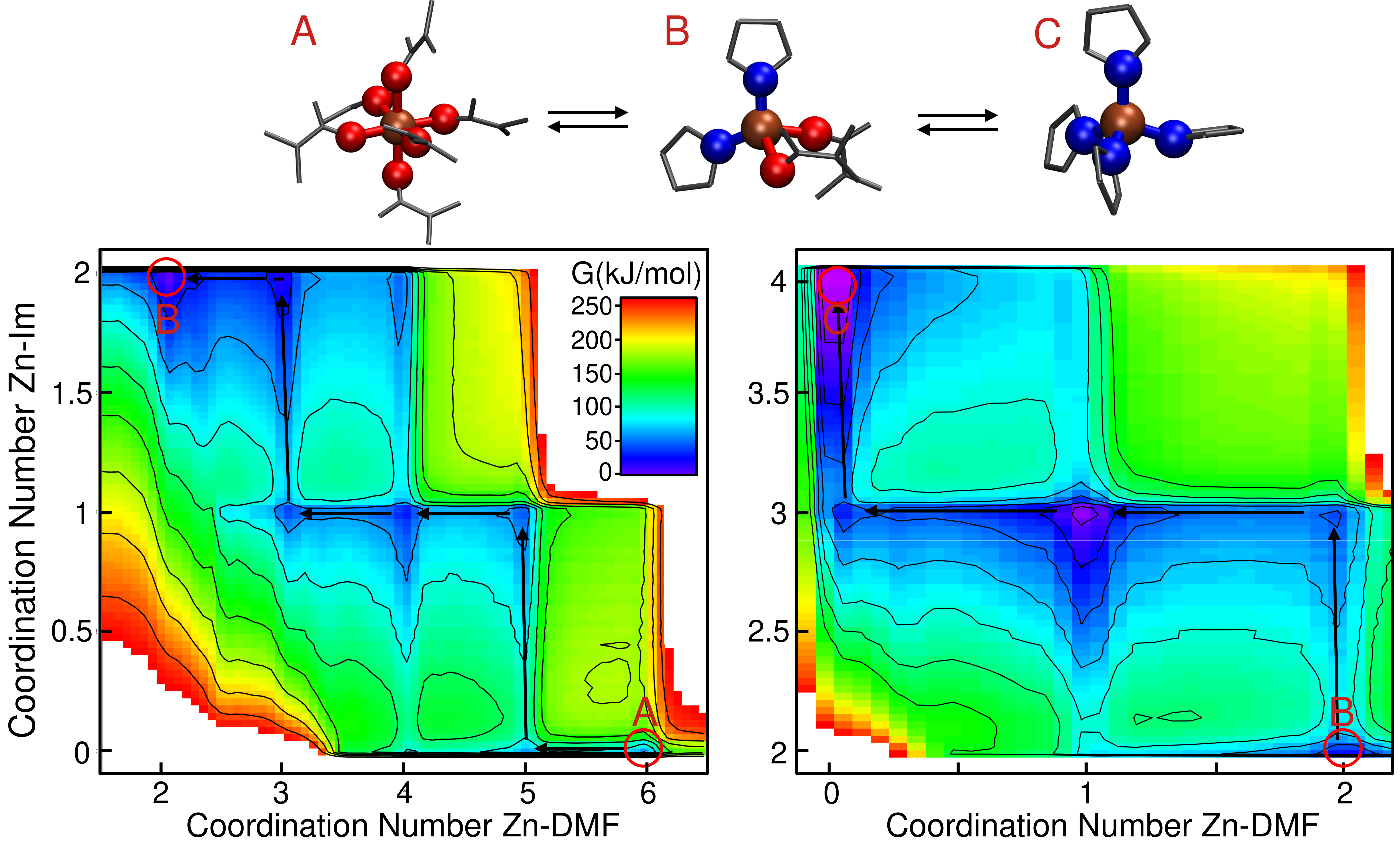}
\caption{\label{fig:2d}{Free energy surface in the space given by $n_{Zn-DMF}$ and $n_{Zn-Im}$ for reactions (I) (left) and (II) (right).
In the left panel, the black arrows indicate the most favorable sequence of steps that connect the fully solvated octahedral complex (state A) with the [Zn(Im)$_2$(DMF)$_2$] tetrahedral complex (state B). In the right panel, the arrows indicate the optimal path for transforming B into the 4-imidazolate-coordinated Zn (state C). The zero of free energy in both plots was located in the absolute minima within each of them. 
Typical configurations of the three states A, B and C are shown in the upper part of the plot. Zn ions 
are displayed in ochre, Zn-bonded O and N atoms are displayed in red and blue respectively. All other species are plotted in gray. H atoms of imidazolate ions are omitted for clarity purposes.}}
\end{figure}

In the left panel of Fig. \ref{fig:2d} we plotted the free energy as a function of the 
Zn-O(DMF) and Zn-N(Im) coordination numbers ($n_{Zn-DMF}$ and $n_{Zn-Im}$ respectively) for the addition of the first two imidazolates. To do so, 
we reduced the dimension of the original four-dimensional free energy surface obtained from metadynamics via a transformation of coordinates explained in the Supporting Information (SI). The zero of free energy was arbitrarily assigned to the global minimum located at $n_{Zn-Im}=n_{Zn-DMF}=2$.
The exact definition of coordination number is given in the Methods Section.

In the lower part of the plot at which $n_{Zn-Im}=0$ we identified three minima that correspond to the [Zn(DMF)$_4$]$^{2+}$ ($G=84\pm3$ kJ/mol), [Zn(DMF)$_5$]$^{2+}$ ($G=60\pm3$ kJ/mol) and [Zn(DMF)$_6$]$^{2+}$ ($G=56\pm4$ kJ/mol) species. 
The [Zn(DMF)$_6$]$^{2+}$ complex is the most stable one as expected.
In the $n_{Zn-Im}=1$ region, we found three other stable complexes: [Zn(Im)(DMF)$_3$]$^{+}$ ($G=38\pm4$ kJ/mol), [Zn(Im)(DMF)$_4$]$^{+}$ ($G=28\pm5$ kJ/mol) and [Zn(Im)(DMF)$_5$]$^{+}$ ($G=43\pm4$ kJ/mol). From these results we can infer that after the addition of the first ligand molecule, the hexacoordinated structures are already destabilized. This can be explained by 
steric effects: the presence of five DMF molecules together with a larger imidazolate ion in the coordination shell of the Zn ion is not favorable. Entropic effects that favor the lower coordinated structures could also be relevant at this high temperature condition.
Finally, when the second ligand is incorporated at $n_{Zn-Im}=2$, the three observed minima are assigned to the [Zn(Im)$_2$(DMF)$_2$] ($G=0\pm6$ kJ/mol), [Zn(Im)$_2$(DMF)$_3$] ($G=15\pm6$ kJ/mol) and [Zn(Im)$_2$(DMF)$_4$] ($G=46\pm5$ kJ/mol) species.
At this point the tetrahedral coordination starts being the most stable.
With this information, we can establish the minimum energy path that connects the states A and B, which is marked with black arrows in Fig.\ref{fig:2d}. The identified mechanism involves the following steps: (i) [Zn(DMF)$_6$]$^{2+}$ loses one solvent molecule, (ii) this allows the incorporation of the first ligand yielding a [Zn(Im)(DMF)$_5$]$^{+}$ complex. (iii) Two consecutive solvent detachment events lead to the first tetracoordinated species: [Zn(Im)(DMF)$_3$]$^{+}$. (iv) Finally, a second ligand is incorporated followed by another solvent loss, giving as result the most stable complex [Zn(Im)$_2$(DMF)$_2$] (state B in the figure).

In the right part of Fig. \ref{fig:2d} we plotted the free energy surface for reaction (II). From this figure we can observe that once the first two imidazolates bind to the Zn ion, the most stable structures for the next additions will retain the tetrahedral geometry.
For the incorporation of the third imidazolate, the new bond is first formed, followed by the loss of a solvent molecule, as it was the case for the second one. For the forth imidazolate addition, the solvent detachment happens before, due to the fact that the high volume of the ligand prevents the possibility of accommodating one DMF in addition of the four imidazolates. 

\subsubsection{Overall reaction free energy profile}

\begin{figure}[h]
\centering
\includegraphics[width=0.5\textwidth]{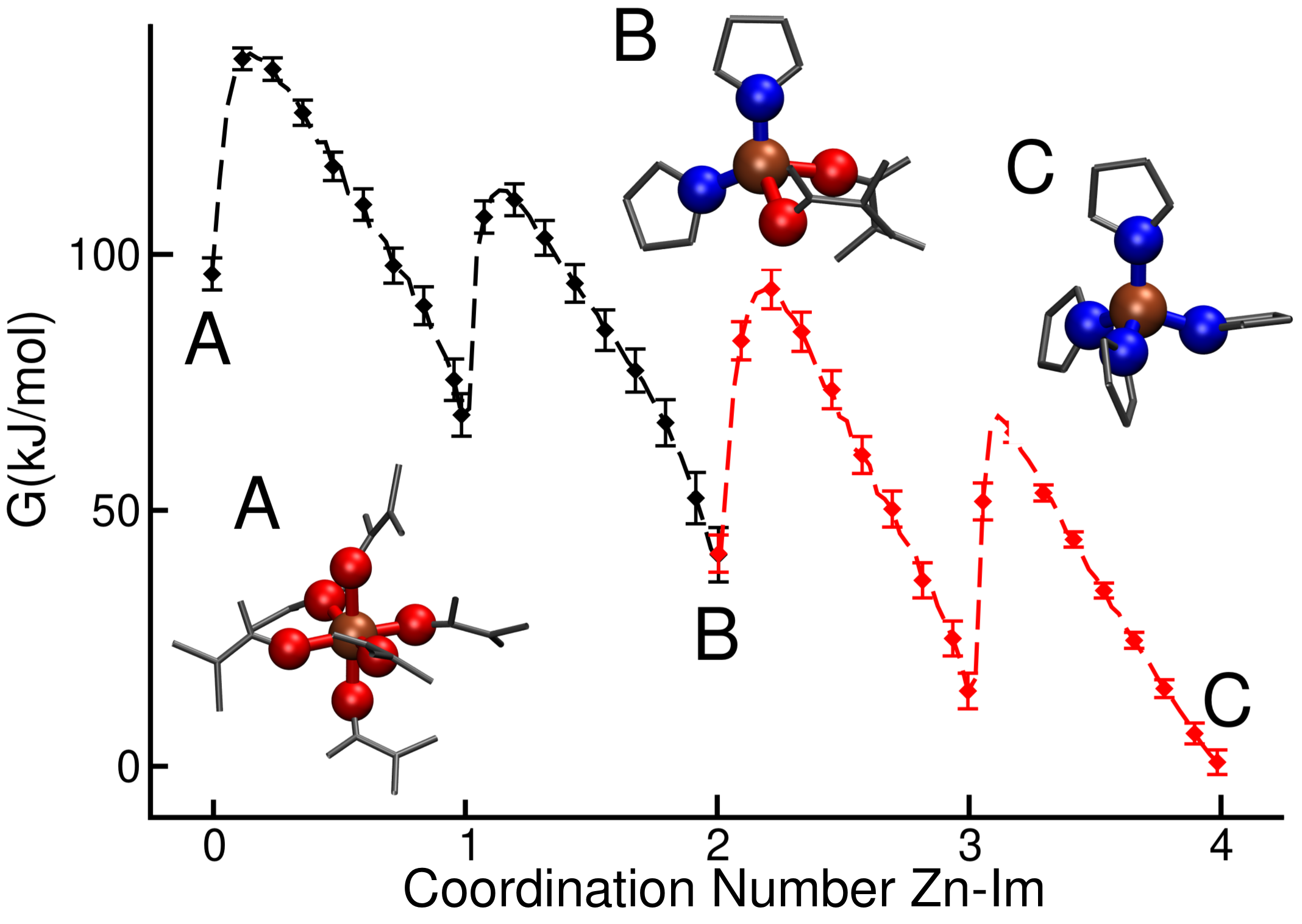}
\caption{\label{fig:znim4}{Free energy as a function of $n_{Zn-Im}$ for the full process involving reactions (I) (black curve) and (II) (red curve). Both curves were aligned to match in the common state (labeled as B). The zero of free energy is arbitrarily assigned to the C species. Representative snapshots of the states A, B and C are also shown.}}
\end{figure}

The free energy landscape for the overall process is 
shown in Fig. \ref{fig:znim4}. In this case $n_{Zn-DMF}$ was integrated out in order to allow a more clear visualization. The black and red curves represent results obtained from the well-tempered metadynamics simulations associated reactions (I) and (II) respectively, and were aligned so that the free energy of the common state, [Zn(Im)$_2$(DMF)$_2$], matches.

From this plot we can see that free energy changes that involve the first, second and third ligand additions are virtually identical ($\Delta G=-28\pm4$ kJ/mol for the first imidazolate incorporation and $\Delta G=-27\pm4$ kJ/mol for the other two). 
We also know from the previous analyses that the first two additions involve the loss of two solvent molecules each, while the third one only involves a single Zn--O(DMF) bond breaking. This last observation would lead us to expect a greater free energy drop for the third 
ligand addition compared to the other two. Nevertheless, this difference in the number of Zn--O(DMF) broken bonds could be compensated by the entropy gain and the lowering of steric repulsion that take place when the total coordination number of the complex is reduced. 
This argument can be tested by analyzing the free energy differences between the [Zn(DMF)$_n$]$^{+2}$ species with $n=4,5,6$ extracted from Fig.\ref{fig:2d}. The free energy drop 
caused by the loss of the first solvent molecule is much lower than the one that corresponds to the second one ($\Delta G_{Zn(DMF)_{6\rightarrow5}}=4$ kJ/mol vs. $\Delta G_{Zn(DMF)_{5\rightarrow4}}=24$ kJ/mol), thus confirming our hypothesis.

Compared to the addition of the first, second and third ligands, the magnitude of the free energy drop that corresponds to the last step is reduced by a half, yielding a value of $\Delta G=-14\pm3$ kJ/mol. This difference can be explained by the steric repulsion between the large imidazolates in [Zn(Im)$_4$]$^{-2}$ as well as the Coulombic repulsion between the negatively charged complex and the newly added ligand.
This trend was also observed by Balestra and coworkers by means of DFT studies for the case of water and ethanol acting as solvent.\cite{Balestra2023}
The binding free energy of the fourth ligand will turn out to be an important feature when comparing this case with the setup studied in next section, in which the tagged Zn starts as part of the surface of a crystalline slab instead of as a dissolved ion.

\subsection{ZIF crystal growth}

In this section, we will tackle the question on whether the energetics of forming a new Zn--N(Im) bond will depend on the self-assembly stage or not. 
In particular, we aim to compare the Zn--Im recombination reactions that occur during the very first nucleation stages with those occurring at the interface of an already formed ZIF crystal slab and an electrolyte solution, which serves as a model of the late growth stages. 

\begin{figure}[h]
\centering
\includegraphics[width=0.9\textwidth]{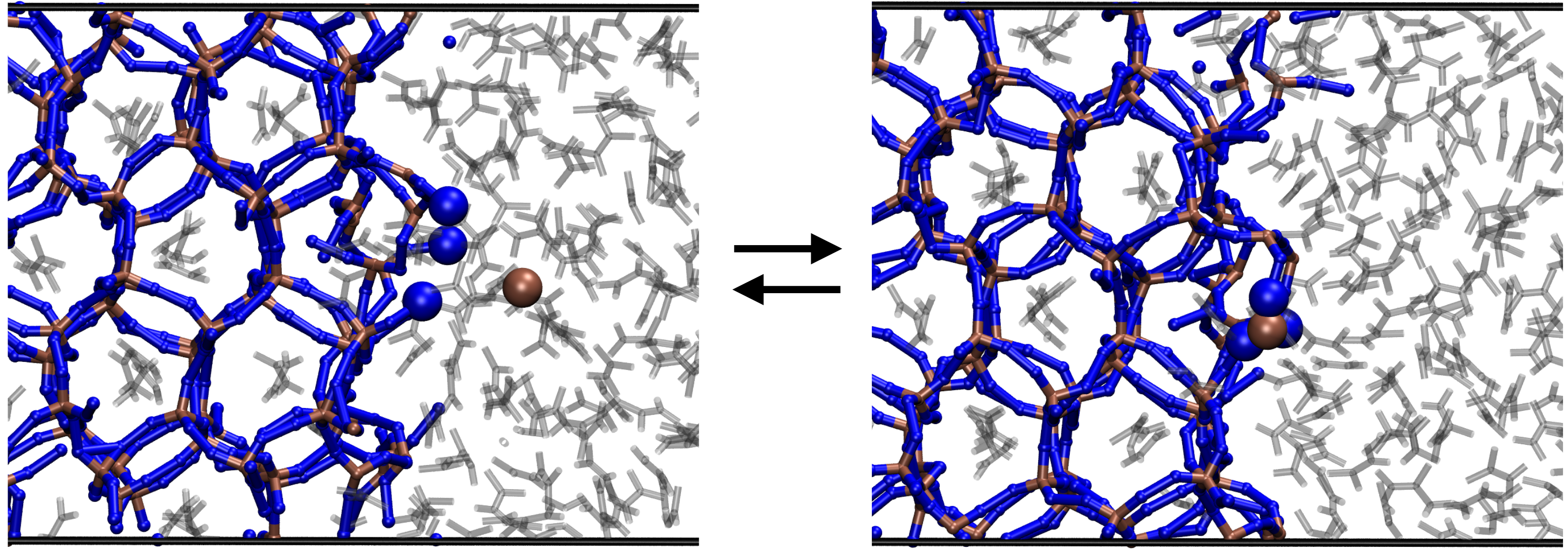}
\caption{\label{fig:adsorption}{Scheme of the model system employed to study the energetics and mechanisms of ZIF crystal growth. The (001) surface of ZIF-4 is shown. The Zn and N atoms involved in the reaction are displayed as spheres. The solvent molecules are plotted in gray. For clarity purposes, only the Zn and N atoms of the crystal slab are shown.}}
\end{figure}

The growth of the ZIF surface will be decomposed into two fundamental reactions: (i) a solvated Zn ion adsorbs into a crystal site composed by three undercoordinated surface ligands (see Fig.\ref{fig:adsorption}) and (ii) a similar process but with the ligand acting as the adsorbed species, this time binding with an undercoordinated surface Zn.

\subsubsection{Zn adsorption-desorption 
on a ZIF surface}

We computed the free energy changes that operate during the reaction shown in Fig.\ref{fig:adsorption} via well-tempered metadynamics simulations, as described above for the study of the very first nucleation stages. 
This time, we used another set of CVs to properly sample the process (see Methods section). 
For quantifying the influence of the surface local structure on the free energy of the reaction, we compared results obtained for four different slabs: three of them correspond to a ZIF-4 crystal but differ in their $(h k l)$ Miller indices. Specifically, the surfaces studied were
cut in the directions perpendicular to the $x$, $y$ and $z$ axes (equivalent to the crystallographic $a$, $b$ and $c$ directions), resulting in Miller indices of (100), (010) and (001) respectively. The forth surface was obtained by cutting a ZIF-1 crystal in a plane perpendicular to the $y$ direction (lattice vector $b$), that corresponds to (010).
The reason behind this choice was to study the effect of changing (i) the Miller indices of the exposed face by comparing results from the three ZIF-4 surface slabs and (ii) the topology of the crystalline structure by comparing results coming from two different polymorphs. ZIF-1 was chosen because it is the second most stable porous polymorph after ZIF-4.\cite{Lewis2009}   
This crystal was cut perpendicular to the $y$ direction to avoid oblique surfaces, given the triclinic nature of the ZIF-1 unit cell. The procedure for generating all the surface slabs is detailed in the SI. In all cases, the binding site for the Zn comprises three free ligand molecules, as shown in Fig. \ref{fig:adsorption}.

\begin{figure}[h]
\centering
\includegraphics[width=0.5\textwidth]{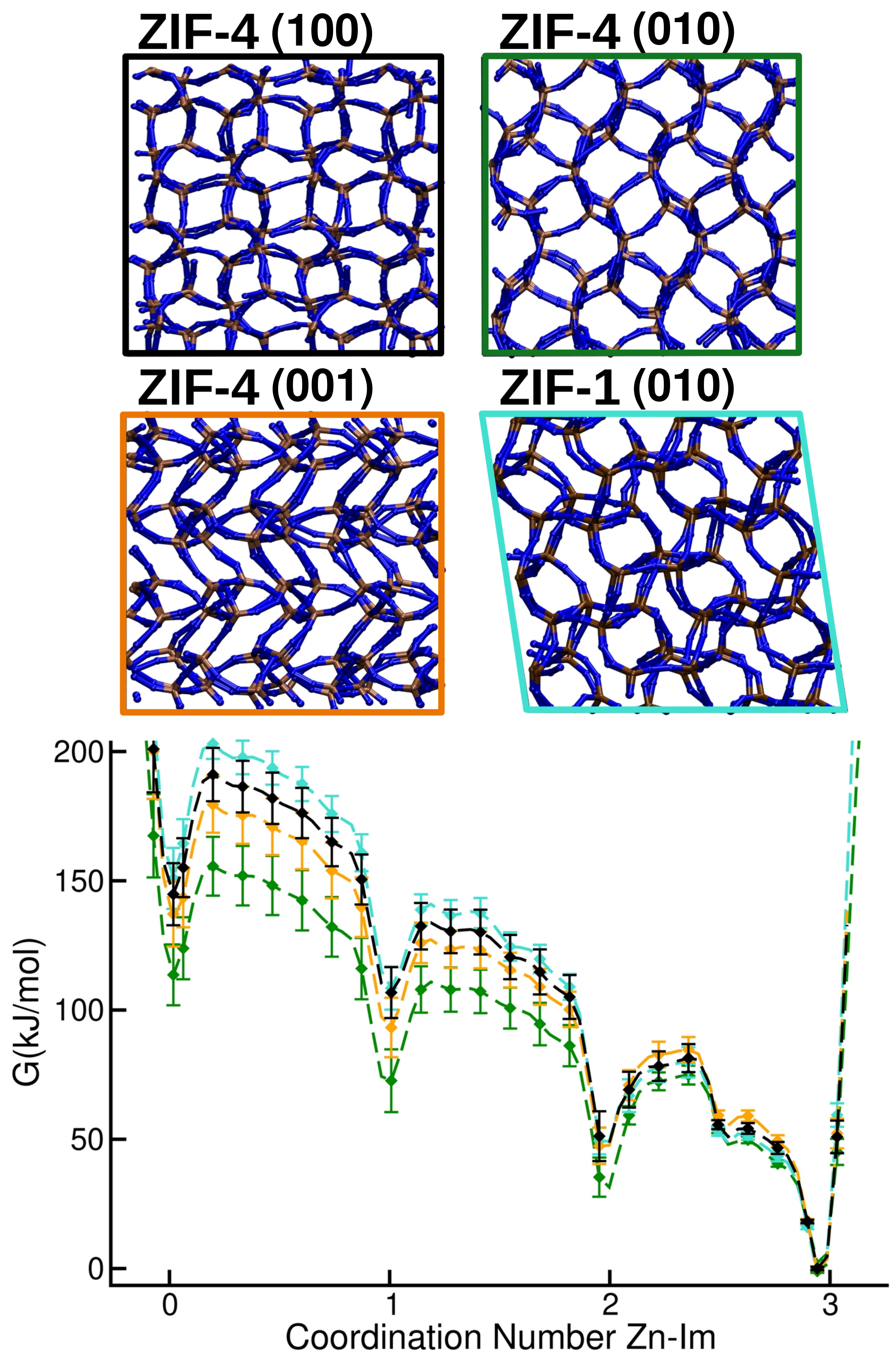}
\caption{\label{fig:zn_add}{Free energy vs. $n_{Zn-Im}$ for the reaction schematized in Fig.\ref{fig:adsorption} for ZIF-4 surfaces with Miller indices (100) (black) (010) (green) and (001) (orange) and for a ZIF-1 (010) surface (turquoise). Images of the surfaces are displayed at the top of the figure.}}
\end{figure}

In Fig. \ref{fig:zn_add} we show the free energy landscapes for these processes as a function of the coordination number between the Zn and imidazolate ions, in the same spirit as in Fig. \ref{fig:znim4}. 
The other two CVs employed in the well-tempered metadynamics procedure were integrated out to make this plot. The zero of free energy was located at the absolute free energy minimum in all cases. 

\begin{table}[h!]
\centering
\small
  \caption{Values of $\Delta G_i$ in kJ/mol for each of the surfaces studied with the corresponding errors. The index $i$ represents the change in the coordination number of the tagged Zn.}
  \label{energies}
  \begin{tabular*}{0.5\textwidth}{@{\extracolsep{\fill}}llll}
     & $\Delta G_{0\rightarrow1}$ & $\Delta G_{1\rightarrow2}$ & $\Delta G_{2\rightarrow3}$\\
    \hline
ZIF-4 (100)      & -38$\pm$11 & -58$\pm$10 & -49$\pm$5  \\
ZIF-4 (010)      & -41$\pm$12  &  -41$\pm$10 & -29$\pm$4   \\
ZIF-4 (001)      &  -41$\pm$12 & -46$\pm$9 & -49$\pm$4   \\
ZIF-1 (010)      &  -42$\pm$10  & -59$\pm$8 &  -51$\pm$4 \\
    \hline
 \end{tabular*}
\end{table}

There are some common features between plots, for example, four minima that correspond to the Zn ion bonded to zero, one, two or three surface ligands are 
always present. The free energy differences between the four minima are shown in Table \ref{energies}. 

An average value of $\Delta G_{0\rightarrow1}=-40$ kJ/mol was registered for the first 
Zn--N(Im) bond formation while for the other two, the observed free energy difference was about $\Delta G_{1\rightarrow2}=$-51 and $\Delta G_{2\rightarrow3}=$-45 kJ/mol on average. This difference is explained by the fact that the Zn has to diffuse from the bulk to the surface vicinity during the first step. All the free energies are higher in absolute value than their counterparts from the [Zn(Im)$_4$]$^{-2}$ complex formation, meaning that the crystal growth is more exergonic than the early nucleation steps, as predicted by nucleation theories from surface tension arguments.\cite{Volmer1926}

All of the free energy differences between consecutive steps lie within the uncertainty bars of the others, meaning that we cannot state that the nature of the surface has an important influence in the Zn adsorption process. An exception to this behavior was found in the plot that corresponds to ZIF-4 (010), which presents a lower energy drop than the other two ZIF-4 surfaces for the step that goes from forming two Zn--N(Im) bonds to three (see $\Delta G_{2\rightarrow3}$).
In order to explain this behavior we can consider the original description of the process in therms of 
$n_{Zn-Im}$ and $n_{Zn-DMF}$.
The mechanism of the third imidazolate bond formation involves, as explained in the 'Formation of early [Zn(Im)$_n$]$^{2-n}$ complexes' section, the bond formation itself followed by a detachment of a DMF molecule from the Zn ion.
We found that the main difference between the (010) surface and the others lies in the step of solvent detachment, which is about $\sim10$ kJ/mol less exergonic in this case.
By visual inspection of the trajectories, we found that this tagged solvent molecule lies inside the ZIF pore in the case of the (010) surface. 
As a consequence, it is harder for this DMF molecule to diffuse far from the Zn than in the other surfaces, in which the solvent molecule to be detached is pointing towards the liquid region. Snapshots of these pentacoordinated intermediate structures for the ZIF-4 (010) and ZIF-4 (001) surfaces are shown in the SI.

\subsubsection{Ligand adsorption-desorption on a ZIF surface}

\begin{figure}[h]
\centering
\includegraphics[width=0.5\textwidth]{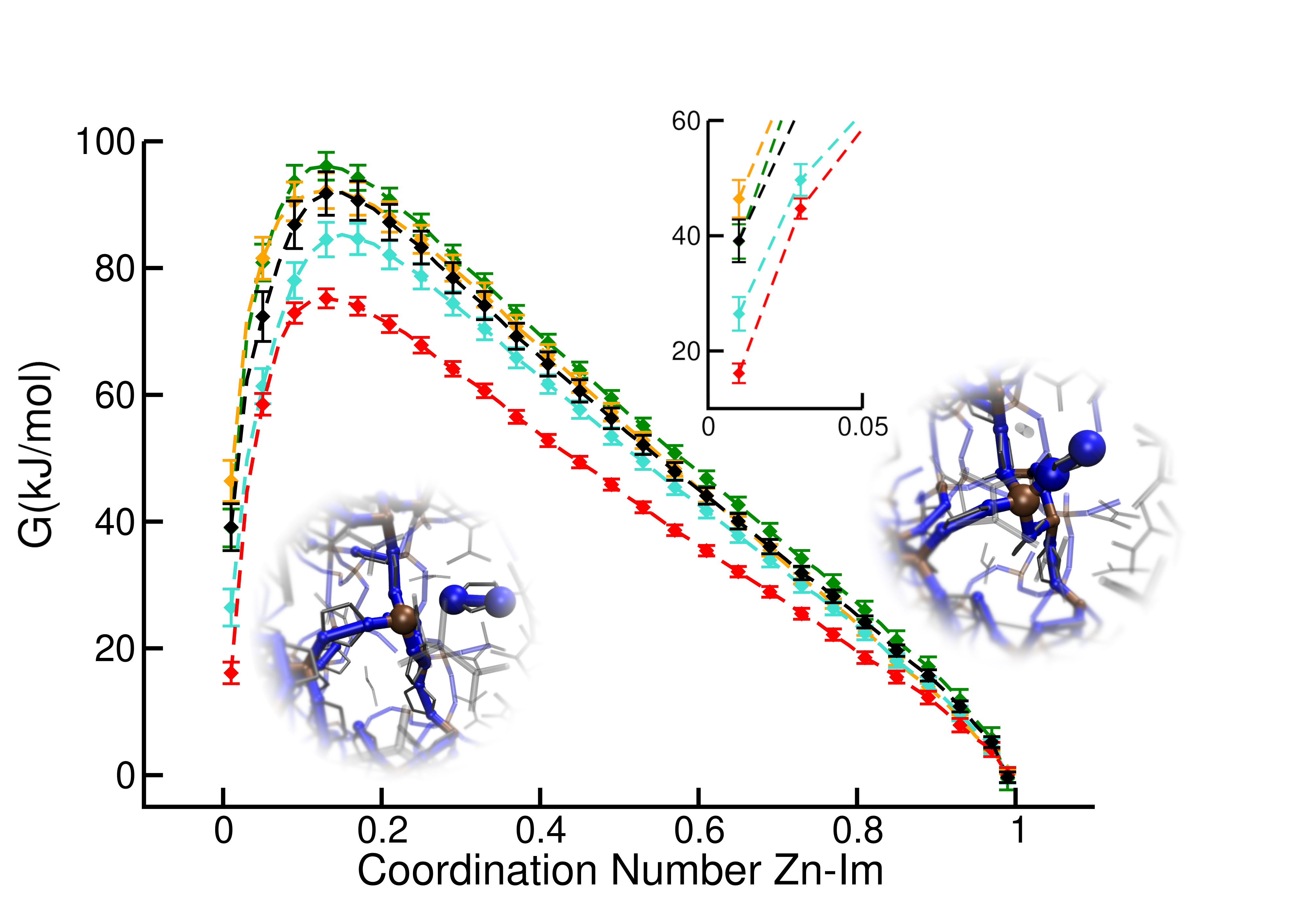}
\caption{\label{fig:lig_add}{Free energy vs. coordination number between the tagged imidazolate and the reactive Zn ion at the ZIF-4 (100) (black), (010) (green) and (001) (orange) and ZIF-1 (010) (turquoise) surfaces and for an isolated [Zn(Im)$_3$]$^-$ complex (red). In the inset we highlight the zero coordination region. Typical snapshots of the non-bonded (left) and bonded (right) species are also shown. }}
\end{figure}

Our analysis concludes with the study of the other adsorption process that takes place during the growth stage of the synthesis: the addition of a ligand molecule to an undercoordinated surface Zn. Since in this case only one bond has to be formed, it was possible to use only two CVs instead of three to properly sample the reaction (see Methods section).
In order to compare with the previous results, we plotted the free energy as a function of the coordination number in Fig. \ref{fig:lig_add}, which in this case goes from zero (non bonded ligand) to one (bonded ligand).
We also include the results for the last ligand addition to the Zn ion in solution that was discussed in section 'Formation of early [Zn(Im)$_n$]$^{2-n}$ complexes' in the figure, for comparison.

For the ZIF-4 surfaces, values of $\Delta G=-41\pm3$ kJ/mol were registered for the (100) and (010) surfaces, while $\Delta G=-47\pm3$ kJ/mol for the (001) one.
The value that corresponds to the ZIF-1 surface slab was $\Delta G=-28\pm3$ kJ/mol and finally for the [Zn(Im)$_3$]$^-$ a value of $\Delta G=-14\pm3$ kJ/mol was found, as stated above.
The addition of the fourth ligand is less energetically favorable than the previous ones in all cases (see Fig.\ref{fig:znim4} and Fig.\ref{fig:zn_add}), because of the already mentioned steric effects, with the exception of the (010) case analyzed above. This suggests that the ligand adsorption could be the limiting step of the process, which would also be in line with the empirical observation made by synthesis experts that an excess concentration of imidazole with respect to the stoichiometric value is required for the synthesis to be successful.\cite{Park2006}
Nevertheless, this difference is much more subtle 
for growth than for early nucleation. The fact that the reactive Zn is already part of the crystal
in the growth stages seems to promote a more adequate geometry for the addition of the last ligand. This makes the formation of the [Zn(Im)$_4$]$^{2-}$ species the least favorable step, which highlights the importance of the stability of 
these initial building blocks in the formation of ZIFs. This stability can be modulated by changes in synthesis conditions, such  as the nature of the solvent, temperature, ligand:metal ratio among others.

With respect to the free energy of binding a ligand for different surface slabs, we found the following trend: ZIF-4(001)>ZIF-4(100)=ZIF-4(010)>ZIF-1(010). 
The lowest value, that corresponds to ZIF-1, can be explained by the fact that
this polymorph is less stable than ZIF-4. This is due to its lower density, which results in the reduction the intensity of van der Waals interactions. \cite{Lewis2009}
In order to explain the difference found between ZIF-4 surface slabs, we characterized the solvation structure of each surface, since  breaking a Zn--O(DMF) bond is the bottleneck of this reaction, as explained above.

\begin{figure}[h]
\centering
\includegraphics[width=0.5\textwidth]{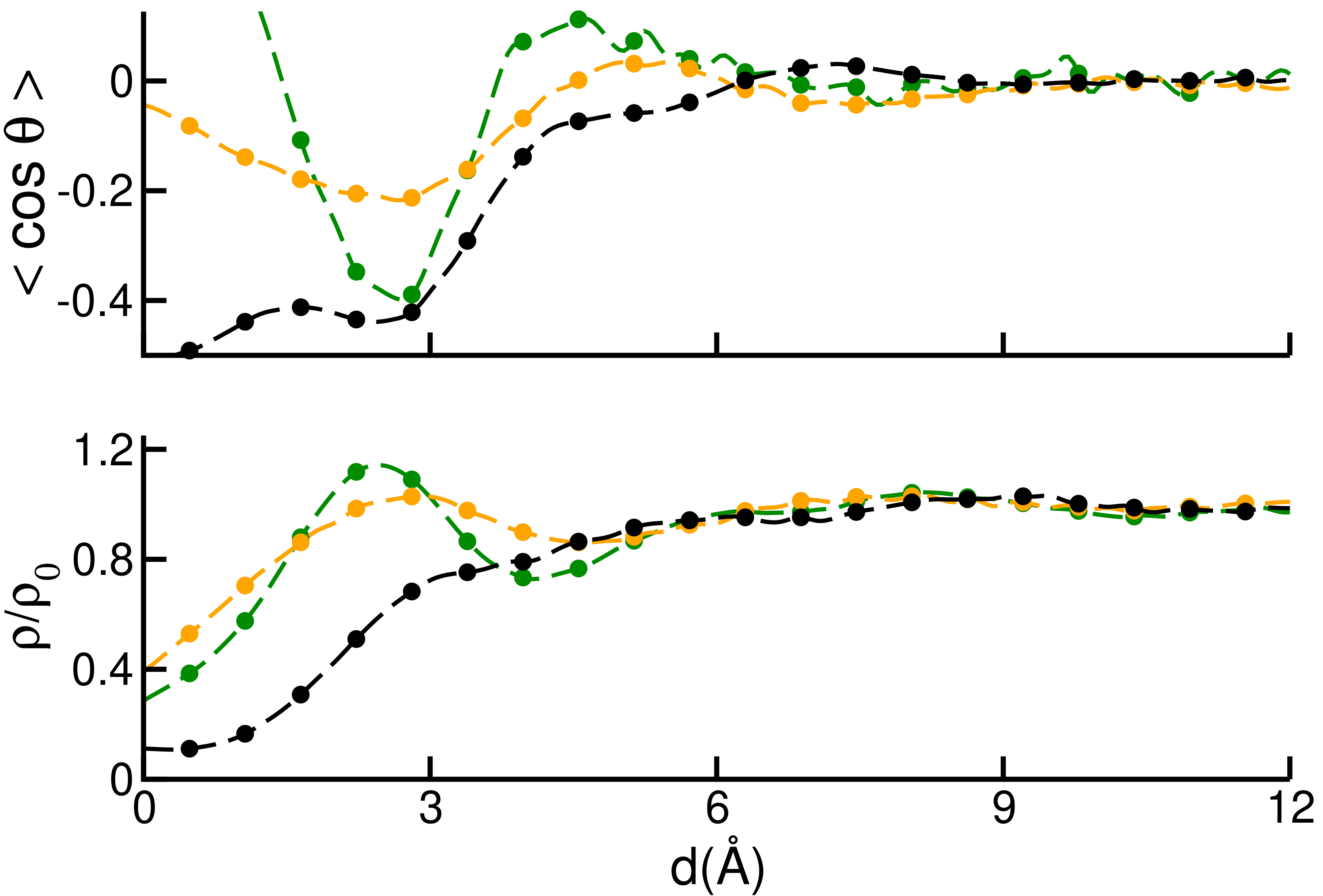}
\caption{\label{fig:orientation}{Average orientation of DMF molecules (top) and average DMF density (bottom) as a function of the distance to the outermost Zn layer of ZIF-4 (100) (black), (010) (green) and (001) (orange) surfaces. Orientations were measured by computing the cosine of the angle $\theta$ between the C(DMF)--O(DMF) bond and the vector perpendicular to the surface plane. Densities were normalized by the bulk value.}}
\end{figure}

In Fig.\ref{fig:orientation}, we plotted the average orientation (top) and density (bottom) of solvent molecules as a function of their distance to the surface. Orientations were measured by computing the cosine of the angle $\theta$ between the C(DMF)--O(DMF) bond and the vector perpendicular to the surface plane. Densities were normalized by the bulk value of $\rho_0=6.80$ nm$^{-3}$. The position of the surface was assigned to the average position of the outermost Zn layer, while that of solvent molecules was described by their central C atom.
From the top panel we observe that in all cases the orientation reaches a minimum at distances around $d\sim3$ \AA, which correspond to the Zn--C(DMF) distance in a bonded Zn--O(DMF) pair, which means that the solvent is oriented perpendicular to the surface in this region. In the (010) and (001) cases, this is accompanied by the presence of a maximum in the density plot (see bottom panel), that comprises the first solvation layer of the surface. For the (100) surface, the density does not reach a maximum at this point, suggesting that the solvent penetrates less the vicinity of the surface slab. Nevertheless, the trend observed in Fig. \ref{fig:lig_add}, seems to 
suggest that the orientation of the solvent molecules plays a more important role than local density in the ligand binding process. For the (100) and (010) surfaces, the average orientation reaches a minimum of -0.4, while for the (001) plane this minimum is reduced to -0.2. This indicates that in the solvent is less strongly aligned to the normal vector of the plane in the case of the (001) surface, which makes it easier to detach. 
This argument can be reinforced by comparing the free energy cost of the DMF remotion from a surface 
Zn, which can be obtained from the same well-tempered metadynamics simulations as the ones that correspond to Fig. \ref{fig:lig_add}.
The obtained DMF detachment free energies were $\Delta G_{(100)}=48\pm3$ kJ/mol, $\Delta G_{(010)}=42\pm3$ kJ/mol and $\Delta G_{(001)}=29\pm3$ kJ/mol for the ZIF-4 (100), (010) and (001) surfaces, respectively. This trend confirms that the DMF is detached more frequently from a (001) surface than from the (100) and (010) ones.
Even though it was found that the ZIF-4 planes present some differences in these aspects, it is clear that the magnitude of the discrepancies in Fig. \ref{fig:lig_add} are not enough to make the ZIF-4 crystals grow faster in $c$ direction than in $a$ and $b$. 
Experimentally this is also confirmed by the fact that the average ZIF-4 crystal shape is not elongated in any preferential direction\cite{Zhang2019}.

\section{Conclusion}

In this work we studied the energetics and mechanisms underlying the early nucleation and late growth stages of the self-assembly of ZIFs by means of well-tempered metadynamics simulations carried out with a reactive force field. We describe an ten-steps mechanism that leads from an octahedral, fully solvated, Zn ion to a tetrahedral complex featuring four ligands:
\small{

(i) [Zn(DMF)$_6$]$^{2+}$ \ce{<=>}[Zn(DMF)$_5$]$^{2+}$ + DMF

(ii) [Zn(DMF)$_5$]$^{2+}$ + Im$^-$ \ce{<=>} [Zn(Im)(DMF)$_5$]$^{+}$

(iii) [Zn(Im)(DMF)$_5$]$^{+}$ \ce{<=>} [Zn(Im)(DMF)$_4$]$^{+}$ + DMF

(iv) [Zn(Im)(DMF)$_4$]$^{+}$ \ce{<=>} [Zn(Im)(DMF)$_3$]$^{+}$ + DMF

(v) [Zn(Im)(DMF)$_3$]$^{+}$ + Im$^-$ \ce{<=>} [Zn(Im)$_2$(DMF)$_3$]

(vi) [Zn(Im)$_2$(DMF)$_3$] \ce{<=>} [Zn(Im)$_2$(DMF)$_2$] + DMF

(vii) [Zn(Im)$_2$(DMF)$_2$] + Im$^-$ \ce{<=>} [Zn(Im)$_3$(DMF)$_2$]$^{-}$

(viii) [Zn(Im)$_3$(DMF)$_2$]$^{-}$ \ce{<=>} [Zn(Im)$_3$(DMF)]$^{-}$ + DMF

(ix) [Zn(Im)$_3$(DMF)]$^{-}$ \ce{<=>} [Zn(Im)$_3$]$^{-}$ + DMF

(x) [Zn(Im)$_3$]$^{-}$ + Im$^-$ \ce{<=>} [Zn(Im)$_4$]$^{2-}$
}

In light of this mechanism, we conclude that the change of coordination number is a complex process involving intermediate pentacoordinated  species. These kind of Zn hybrid complexes has been already characterized both experimentally and theoretically\cite{AdurizArrizabalaga2023,Pucheta2016,Sola1991,Piskunov2017,Kirchner1987}.

The incorporation of the first three ligands is equally favorable in terms of free energy in the early nucleation steps. During the late growth, however, it is less favorable to form the first Zn--N(Im) bond than forming the next two. This is associated to the local environment of the ZIF surface, which offers easy access to undercoordinated ligands. Conversely, adding a fourth ligand is the least exergonic step in the process in early nucleation. When this reaction is compared with its analogous of forming a Zn--N(Im) bond between a tri-coordinated surface Zn and a free, solvated, Im ion, the latter is more exergonic. This suggests that the formation of the fully-coordinated Zn ion is a crucial part in the nucleation. This process can be modulated by changing synthesis conditions, including nature of solvent and reactants, temperature and Zn:Im and reactant:solvent ratios.  

The free energy associated to the addition of a fourth ligand to a surface Zn, a late growth step, depends on the local environment surrounding the tagged Zn. This effect is more important when comparing the growth of two polymorphs than when comparing the growth of ZIF-4 crystal faces with different Miller indices. The differences found in this latter case are associated with the ease of removing the solvent molecules that bind to the surface Zn. Even more subtle differences take place when adding a Zn to a ZIF surface. These are also associated to the removal of solvent molecules and seem to be related to whether the solvent molecules competing with the Zn lie in the solution or whether they are adsorbed in the outermost pore layer of the surface.

This work sheds light into important thermodynamic and mechanistic aspects of the self-assembly of ZIFs. The intermediate nucleation steps that were not covered in this study will be the object of further studies. Our methodology can be extended to study the self-assembly processes of other MOFs and thus contribute to coming closer to the holy grial of MOF rational design.   

\section{Methods}

\subsection{Simulation setup}

All the simulations were performed using the LAMMPS open source software\cite{lammps} coupled with the PLUMED package\cite{Tribello2014}.
The Zn and Im ions were modeled via the nb-ZIF-FF force field\cite{Balestra2022}, in which the Zn--N interactions are represented by a Morse potential that allows bond formation and breaking events.
This force field also incorporates dummy atoms attached to the Zn and N species to correctly reproduce the angular distribution of ligands around a Zn centre.
nb-ZIF-FF adequately reproduces the properties of several ZIF-4 polymorphs, including those that result from its thermal amorphization and melting\cite{Mendez2024} (ZIF-amorphous and ZIF-liquid), its high pressure phases\cite{Mendez2024_2} (ZIF-4-cp and ZIF-4-cp-II) and the ZIF-1 crystal.\cite{Balestra2022}
This force field was also implemented for the study of the ZIF-8 self-assembly process starting from Zn and 
2-methylimidazole ions dissolved in methanol\cite{Balestra2022}. 
For the DMF solvent we employed a flexible version of the CS2 potential developed by Chalaris and Samios\cite{Chalaris2000}. The intramolecular bond, angular and dihedral parameters
were taken from the CHARMM general force field\cite{Vanommeslaeghe2009}.
In order to model the Zn--N(Im) and Zn--O(DMF) interactions on an equal footing, we replaced the Lennard-Jones term of this last pair for a Morse-like potential.
Since the original nb-ZIF-FF was not parameterized 
considering the interaction with DMF, we had to re-optimize simultaneously the Morse depth parameters of both Zn-O(DMF) and Zn-N(Im) pairs. 
We aimed to optimize the force field to reproduce two qualitative aspects of the system observed experimentally: (i) the most stable Zn$^{2+}$ complex in DMF should be octahedral\cite{Ishiguro1990} and (ii) a ZIF-4 crystal filled with DMF should be stable at 
400 K, which is the synthesis temperature\cite{Park2006}.
The final parameters of the force field are summarized in the SI.

All the results were obtained from simulations 
made at the NPT ensemble at $T=400$ K and
$P=1$ bar, which reproduces the experimental solvotermal synthesis conditions of ZIF-4\cite{Park2006}. The time step was set to 0.5 fs. 
Nosé-Hoover thermostats and barostats were used with damping times of 100x and 1000x time steps respectively. 

\subsection{Metadynamics}

Well-tempered metadynamics\cite{Barducci2008} enables the exploration of a selected CV space in a reversible fashion. 
As a result, it is possible to recover the free energy surface as a function of the CV values when convergence is reached.
The selection of 
CVs has to be done in such a way that the two or more states that are being compared are located at different points in this reduced coordinate representation.
This necessary condition is often not sufficient to force the system to visit all the desired states. It is also mandatory that the selected reaction coordinates capture the slowest
or activated degrees of freedom of the reaction.

We chose three CVs for the simulation of the solvated Zn ion that interchanges two ligand molecules shown in the 'Formation of early [Zn(Im)$_n$]$^{2-n}$ complexes' section. The first two of them correspond to the minimum distance between the 
Zn and the two N atoms ($d_{min}$=min\{$d_{N_1}$,$d_{N_2}$\}) of each imidazolate that is being interchanged.
In this way, the Zn may bind with any of the available sites of each of the two ligands.
The functional form of the minimum distance $d_{min}^{(i)}$ to the $i$-th ligand is selected so that it has continuous derivatives by using the following formula:
\begin{equation} \label{mindist}
    d_{min}^{(i)} = (d_{N^{(i)}_1}^{-6}+d_{N^{(i)}_2}^{-6})^{-1/6} \hspace{1cm} i=1,2
\end{equation}
Since the bottleneck of the reaction is the dissociation of a DMF molecule from the solvation shell of the Zn ion, the third CV was chosen to be the coordination number between Zn and the O atoms of the solvent ($n_{Zn-O}$), which was calculated by the following sum over all the O atoms 
($O_i$):
\begin{equation} \label{coordnum}
    n_{Zn-O} = \sum_i f(r_i)
\end{equation}
Where $f(r_i)$ is a function that takes the value of 1 if the Zn-O$_i$ distance $r_i$ is lower than a threshold value of $d_0=2.1$\ \AA\ similar to the bond length and decays to zero afterwards with a characteristic width of $r_0=0.5$\ \AA. In this case:
\begin{equation} \label{activation}
    f(r) = \frac{1-\left(\frac{r-d_0}{r_0}\right)^3}{1-\left(\frac{r-d_0}{r_0}\right)^6}
\end{equation}

For the second stage of ligand additions, the same set of CVs was used, in this case the Zn-ligand distances correspond to the two new molecules that are interchanged, while the previous two imidazolates are kept connected to the Zn through a harmonic constraint. An extra Zn ion is included in the box to maintain the electroneutrality of the system, with a constraint that ensures that it always remains far away from the reactive sites.
The process of going from a free Zn to a fully-coordinated Zn(Im)$_4^{2-}$ was divided in two steps because performing metadynamics simulations with more than three CVs is not recommended due to convergence problems.

For the Zn adsorption-desorption process explained in section 'ZIF crystal growth' a slightly different set of CVs 
was used. Since it was not possible to control the three Zn-N distances 
and the Zn-O coordination number at the same time due to the 
low scaling convergence of metadynamics, we employed the following reaction coordinates: (i) the coordination number between Zn and oxygen, (ii) the coordination number between Zn and nitrogen, and (iii) the distance between the Zn and the center of mass of the three reactive nitrogen sites. This last CV was introduced to probe the slow diffusive motion of the Zn ion from the vicinity of the crystal slab to the bulk region, given that the second CV 
takes the value of zero immediately after the Zn abandons the surface vicinity.

Finally, for the study of the ligand-surface interchange, the lack of multiple reactive sites allowed us to employ only two CVs: (i) the coordination number between the reactive Zn and the O(DMF) atoms and (ii) the minimum distance between the the ligand nitrogen sites and the reactive Zn, in the same fashion as explained in Eq. \eqref{mindist}.
Again, the connectivity between the mobile ligand and any other surface Zn was kept fixed and equal to zero and a upper boundary of 10 \AA\ was imposed to the distance CV to avoid exploring spurious regions of the free energy surface.

For the visualization of the obtained free energy surfaces it is often necessary to perform a dimensionality reduction over the full CV space, keeping either one or two relevant coordinates or a function of them. The procedure used to compute these transformations, as well as further details of the metadynamics simulations and convergence criteria, are explained in the SI.

\begin{acknowledgement}

This work was funded by the European Union ERC Starting grant MAGNIFY, grant number 101042514. This work was granted access to the HPC resources of IDRIS under the allocation
A0150911989 made by GENCI.
\end{acknowledgement}

\newpage

\section{
        Supporting Information for: Thermodynamic Insights into the Self-assembly of Zeolitic Imidazolate Frameworks from Computer Simulations
    }
    {
        Emilio Méndez and Rocio Semino           
    }{
    \newline
    Sorbonne Université, CNRS, Physico-chimie des Electrolytes et Nanosystèmes Interfaciaux, PHENIX, F-75005 Paris, France
    }

\subsection{Well-tempered metadynamics Setup}

The parameters for the well-tempered metadynamics simulations were set as follows:

\begin{itemize}
    \item the initial gaussian height was set equal to $kT$,
    \item the gaussian widths were 0.1 for the coordination number collective variables (CVs) and 0.5 \AA\ for the distance CVs, and
    \item the bias factor that controls the decay of the heights with time was chosen to be 30.
\end{itemize}

Five parallel walkers were employed to accelerate the convergence.\cite{Raiteri2005} Each walker evolves independently from the others but they all share the same bias potential obtained from the addition of gaussian terms.
The total time for each simulation, comprising all the walkers, was around 150 ns. 

In the case of the simulations that involve a crystal slab, the coordination number between the reactive Zn$^{2+}$/ligand and all the free surface N/Zn other than the ones that belong to the tagged site was kept 
fixed and equal to zero. Upper and lower boundaries were imposed to the CVs to avoid the exploration of non-physical regions and to keep the Zn$^{2+}$ within a distance lower than 10 \AA\ from the surface, at which the free energy already 
reaches a plateau.

\subsection{Convergence and uncertainty calculation}

\begin{figure}[h]
\centering
\includegraphics[width=0.55\textwidth]{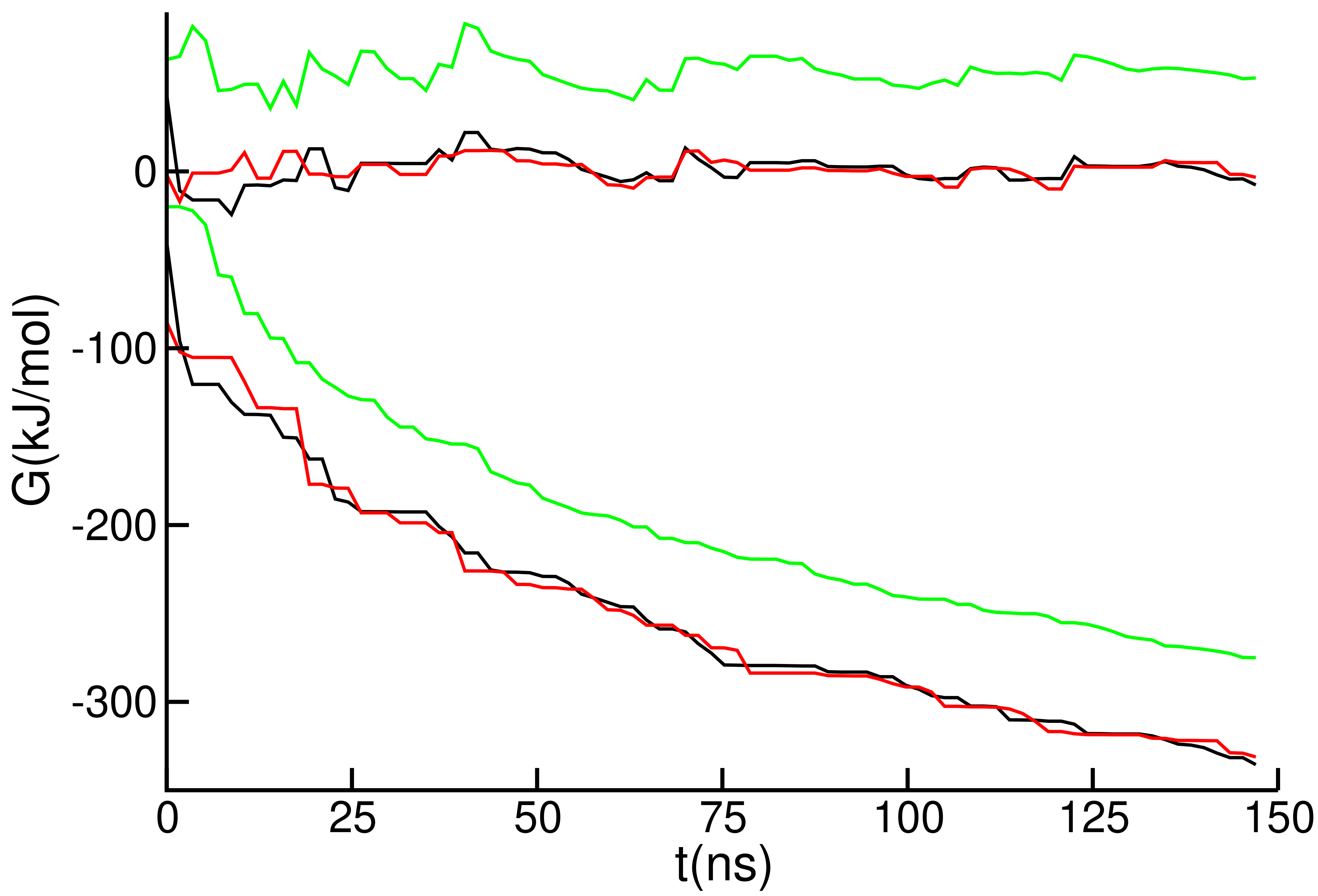}
\caption{\label{fig:estimator}{Free energy estimator for three representative points in the CV space for the reaction that involves the bonding of two imidazolates to a Zn ion. The curves in the negative region correspond to the estimator without the second term of Eq. \ref{free}. The black, red and green curves correspond to the three points described in the text, respectively.}}
\end{figure}

To analyze the convergence of the well-tempered metadynamics simulations we followed the procedure developed by Tiwary \textit{et al.}\cite{Tiwary2014}
This approach takes into account the fact that 
the bias potential is dynamically modified as the metadynamics simulation advances and it never reaches a plateau value. 
This makes it non trivial to choose a convergence criterion.
Tiwary and coworkers found a way to compute a time independent free energy estimator that allows to compare results measured at different times during the simulation given by:

\begin{equation} \label{free}
    G(s) = -\frac{\gamma V(s,t)}{(\gamma -1) } + k_b T \; \mathrm{ln} \int  e^{\frac{\gamma V(s,t)}{(\gamma -1) k_b T}} \; ds 
\end{equation}
were $s$ represents the collective variable(s), $\gamma$ the bias factor, and $V(s,t)$ the time dependent bias potential. 
The last term of Eq. \ref{free} is a time dependent constant that aligns the free energy estimation at time $t$ with the ones computed at previous times. 
In order to apply this technique to data obtained from different walkers, we ordered the gaussians coming from each simulation as a function of time.
As an example, in Fig. \ref{fig:estimator} we plotted the free energy estimator of Eq. \ref{free} as a function of time for three points in the CV space for the reaction that involves the bonding of two imidazolate ions to a Zn ion. The points correspond to (i) d$_1$=2 \AA, d$_2$=6 \AA, $n_{Zn-O}$=4 (ii) d$_1$=6 \AA, d$_2$=2 \AA, $n_{Zn-O}$=4 and (iii) d$_1$=6 \AA, d$_2$=6 \AA, $n_{Zn-O}$=6, where d$_1$ and d$_2$ are the distances between the Zn and the tagged imidazolate moieties and $n_{Zn-O}$ is the number of solvent molecules surrounding the Zn ion. The first two points are 
equivalent and correspond to a situation where one of the imidazolate ions is bonded to the Zn and other one is not. The fact that both curves lie close to each other is another indicator of the convergence of the simulation. The third point is higher in energy and corresponds to a situation where both ligands are dissociated from the Zn ion.
We also plotted the free energy without the addition of the second term of Eq. \ref{free}.
As expected, these last values continue to descend without reaching a plateau, but the corrected estimators fluctuate around constant values.

\begin{figure}[h]
\centering
\includegraphics[width=0.6\textwidth]{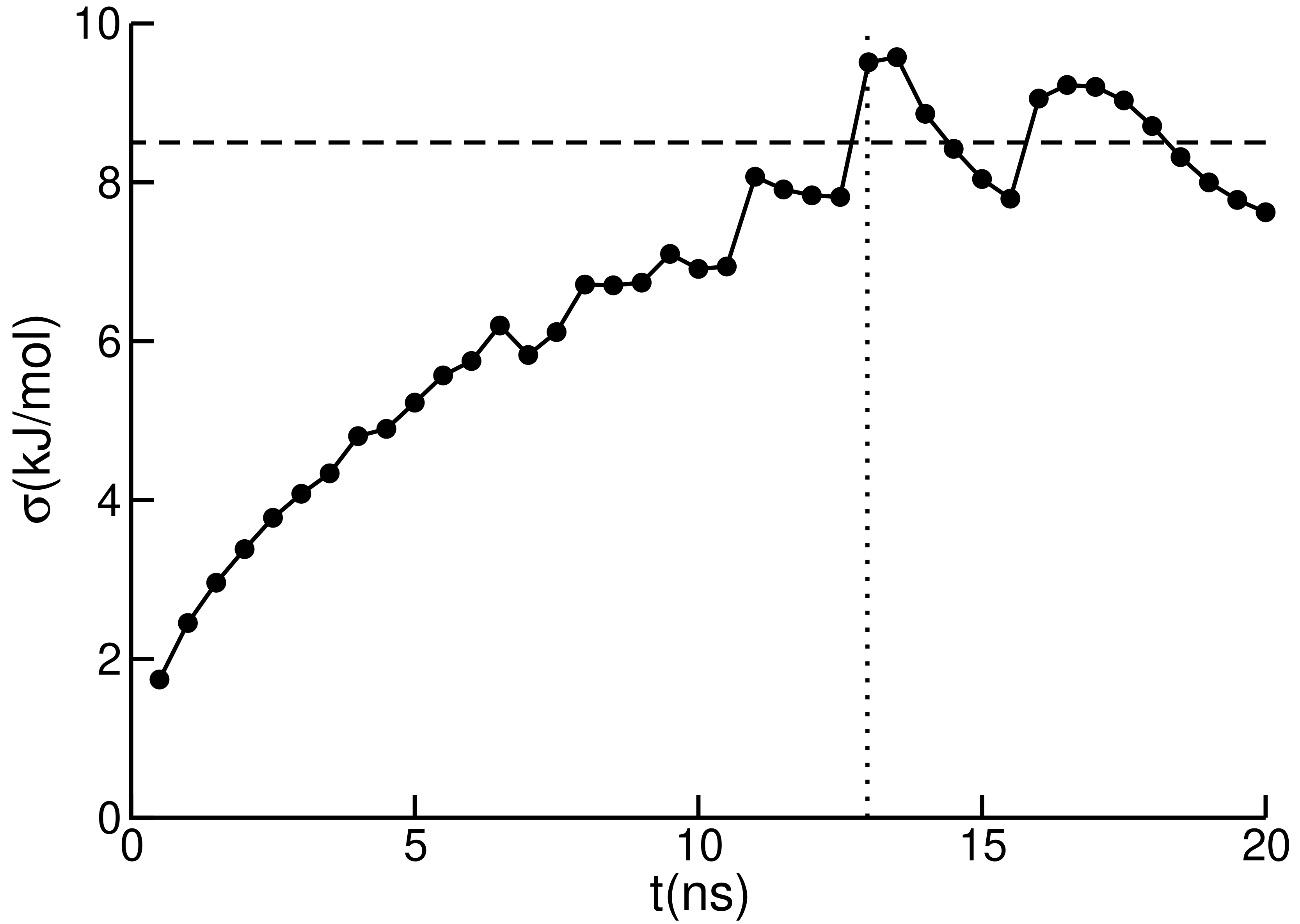}
\caption{\label{fig:blocks}{
Standard deviation of the free energy associated to the absolute minimum as a function of the block size, for the same reaction as 
in Fig. \ref{fig:estimator}. The vertical line indicates the value at which we consider the results to be uncorrelated. The horizontal line indicates the average standard deviation after the decorrelation time is reached.}}
\end{figure}

In order to compute the final free energy profile and the corresponding errors, we need to time average the results from the corrected free energy curves. To avoid artifacts that arise when dealing with correlated data, we employed the block averaging technique developed by Bussi and Tribello.\cite{Bussi2019}
To estimate the optimal block size for which the data is uncorrelated, we computed the standard deviation of the free energy as a function of the block size. 
Results associated to the lowest energy structure are shown in figure \ref{fig:blocks}. When the individual block values become uncorrelated, the standard deviation reaches a plateau. According to this criterion, we averaged data from blocks of 13 ns.
This procedure was performed for all the reactions studied.

\subsection{Transformations of the collective variable space}

It was often necessary to perform some kind of coordinate transformation or dimensionality reduction to have a clearer visualization of the free energy curves.  
All the modifications applied rely on the relationship between the free energy $G$ as a function of the CVs ($\xi_1$, $\xi_2$ and $\xi_3$) and the probability distribution function $\mathcal{P}(\xi_1,\xi_2,\xi_3)$: 

\begin{equation} \label{prob}
    \mathcal{P}(\xi_1,\xi_2,\xi_3) = Ce^{-\beta G(\xi_1,\xi_2,\xi_3)}
\end{equation}
where $C$ is a normalization constant. The undesired CVs can be integrated from the probability distribution function in order to reduce the dimensionality of the free energy surface. Then, the free energy in the reduced space is obtained by inverting Eq. \ref{prob}. 

In the first section of the results we applied the following transformation for computing $G$ as a function of the Zn-Im coordination number ($n_{Zn-N}^{*}$) from $G(d_1,d_2)$. We computed the probability of the new variable by:

\begin{equation} \label{int}
    \mathcal{P}(n_{Zn-N}^{*}) = \int \delta(n_{Zn-N}^{*}-n_{Zn-N}(d_1,d_2)) \; \mathcal{P}(d_1,d_2)\;\mathrm{d}d_1\;\mathrm{d}d_2
\end{equation}
where the function $n_{Zn-N}(d_1,d_2)$ follows the definition given in the main text. In practice, this integration is performed numerically by discretizing the CV space into finite bins. 
Subsequently, the free energy was recovered by the inversion of Eq. \ref{prob}.
The uncertainties of the transformed free energies were 
computed from the ones that correspond to the original curves by propagation of errors.

\subsection{ZIF surface generation}

In order to create the crystal-solvent interfaces described in section 'ZIF crystal growth' of the article we proceeded as follows: 
\begin{enumerate}[label=(\roman*)]
    \item Starting from a ZIF-4 unit cell we filled the system with solvent using a grand canonical Monte Carlo (GCMC) procedure until the experimental loading of 8 DMF molecules 
was reached\cite{Bennett2011_3}. 
    \item The system was then replicated twice in each direction parallel to the desired surface and four times in the direction perpendicular to it.
For example, to construct a surface slab with a normal in the $z$ direction, we should multiply the original unit cell by 2x2x4.
    \item In order to cut the system and generate the interface, we deleted all the atoms that lied outside the central 2x2x2 region. Imidazolate moieties that were
    half-cut during this procedure were
completely removed.
    \item We randomly deleted some of the surface Zn, taking care that both surfaces have the same amount of exposed Zn and ligands. This was done so that the net dipole of the final structure in the direction perpendicular to the surface was zero.
    \item The empty space generated after cutting the MOF, which
    occupies half of the simulation box, was filled with solvent via
    GCMC simulations as done before. 
    The central surface slab was kept frozen during this step.
    \item 
    To avoid any further degradation in the surface other than the desired reaction, we forced the Zn-imidazolate connectivity to remain unaltered in all cases
except for that of the tagged Zn ion or ligand that will be 
adsorbed/desorbed into the surface. This was done by adding extra harmonic 
bonds between neighbor Zn and N atoms. These constraints do not produce any significant structural change in the crystalline slab.
    \item  A short preliminary run of $\sim1$ns was performed to allow the system to equilibrate.
\end{enumerate}

This scheme was also applied for the generation of the ZIF-1 slab. In this case, given the lack of experimental information about the solvent filled structure, we added 24 DMF molecules per unit cell. This corresponds to one molecule per pore in the system, which is equivalent to what was found for ZIF-4, and seems to represent the most stable configuration obtained via GCMC.

\newpage

\subsection{Pentacoordinated intermediate species for the addition of a third ligand to (010) and (001) ZIF-4 surfaces}

\begin{figure}[h]
\centering
\includegraphics[width=0.7\textwidth]{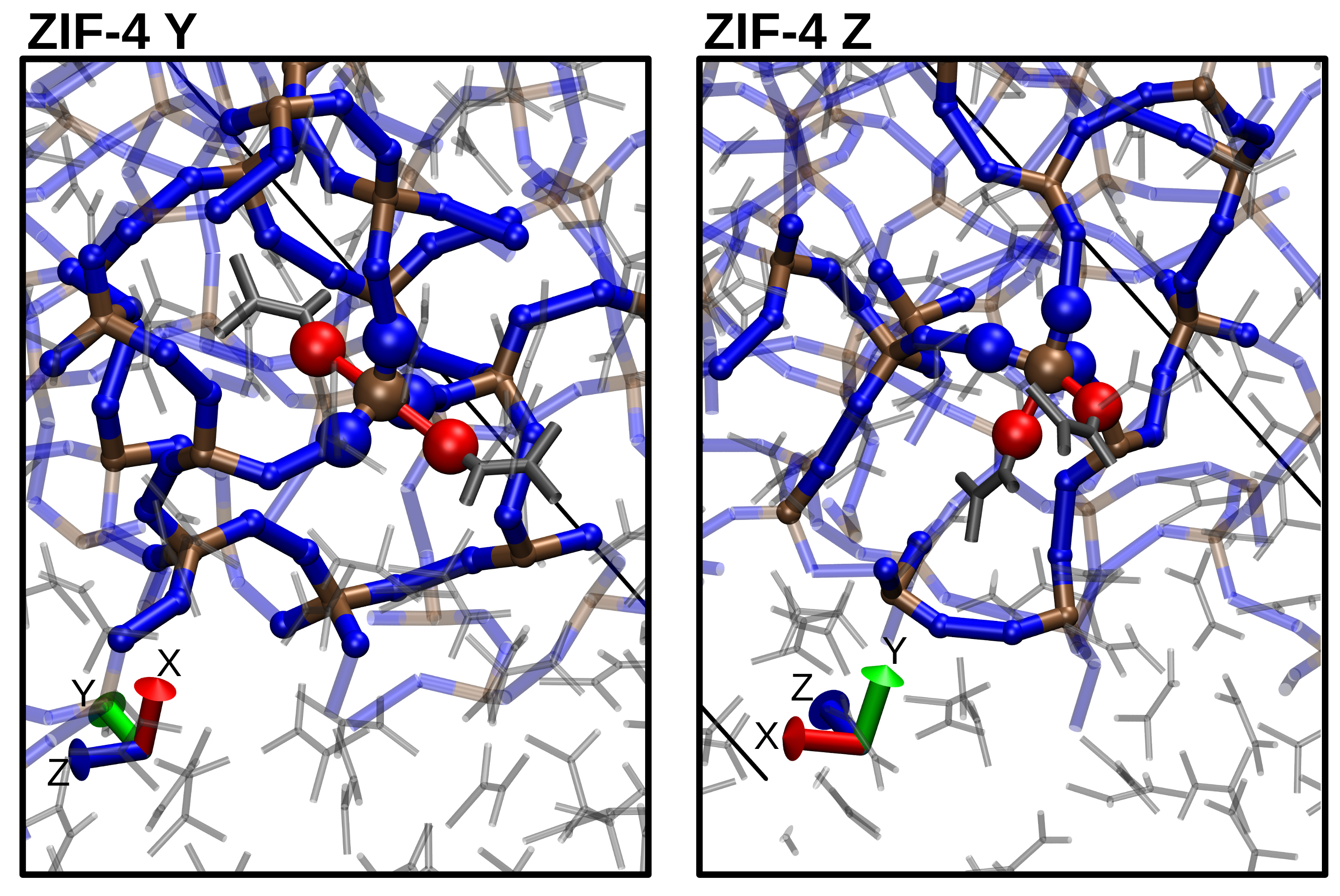}
\caption{\label{fig:penta}{Typical snapshots of the intermediate species before the formation of the tricoodrinated Zn-imidazolate complex in the (010) and (001) ZIF-4 surface slabs (left and right respectively).}}
\end{figure}

\subsection{Force field parameters}

\begin{figure}[h]
\centering
\includegraphics[width=0.22\textwidth]{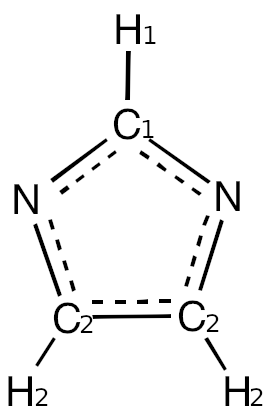}
\caption{\label{fig:imidazole}{
Representation of an imidazolate ion with the atom type name for each species.}}
\end{figure}

The potential energy of the system $E$ was described as a sum of the following contributions:

\begin{equation} \label{lj}
    E = E_{coul-LJ} + E_{morse} + E_{bond} + E_{angle} + E_{dihedral} + E_{improper}
\end{equation}
where $E_{coul-LJ}$ is the coulombic plus Lennard Jones energy and $E_{morse}$ is the Morse energy, and together they constitute the non bonded interactions.
$E_{bond}$, $E_{angle}$,$E_{dihedral}$ and $E_{improper}$ are the intramolecular contributions and refer to the bond, angular, dihedral and improper energies respectively. The formula employed for the calculation of $E_{coul-LJ}$ is the following:

\begin{equation} \label{lj}
    E_{coul-LJ} = 4\epsilon \left[ \left( \frac{\sigma}{r} \right)^{12} - \left( \frac{\sigma}{r} \right)^{6}  \right] + \frac{C q_i q_j}{r} 
\end{equation}
where $r$ is the interatomic distance, $\epsilon$, $\sigma$ and $q$ are parameters that depend on the atomic species while $C$ is a constant. $\epsilon$ and $\sigma$ were obtained from single atom values by standard Lorentz-Berthelot mixing rules, with the exception of Zn--N(Im) and Zn--O(DMF) pairs that do not have Lennard Jones parameters since their interactions are modeled with Morse potentials. In figure \ref{fig:imidazole} we indicate the name of each species in the imidazolate ion (Im).
In table \ref{long} we summarize the $\epsilon$, $\sigma$ and $q$ values for all the present species. The species marked with a $^*$ symbol represent the dummy atoms, that are present in the Zn and N(Im) species. A six sites model was used for dimethylformamide (DMF), in which the methyl groups were considered as united atoms\cite{Chalaris2000}. Long range coulombic interactions were computed with the particle-particle particle-mesh method. The cutoff distance for other interactions was set to 13 \AA. Non bonded interactions were not considered for first and second neighbor atoms, and were scaled by a factor of 0.5 (0.6874) for Lennard Jones (Coulombic) interactions. The Morse potential was computed using the following expression:

\begin{equation} \label{morse}
   E_{morse} = D_0\left[ e^{-2\alpha(r-r_0)} - 2 e^{-\alpha(r-r_0)}  \right]
\end{equation}

where $D_0$, $\alpha$ and $r_0$ are parameters that are also displayed in table \ref{long}. Only Zn--N(Im) and Zn--O(DMF) pairs contributed to the total energy with this kind of interaction. For the bonded terms the employed formulas are:

\begin{equation} \label{bond}
    E_{bond} = K(r-r_0)^2
\end{equation}
\begin{equation} \label{angle}
    E_{angle} = K(\theta-\theta_0)^2 + K_{ub}(r-r_{ub})^2
\end{equation}
\begin{equation} \label{dihedral}
    E_{dihedral} = K[1+\text{cos}(n\phi-d)]
\end{equation}
\begin{equation} \label{improper}
    E_{improper} = K[1+d \text{cos}(n\phi)]
\end{equation}
For these expressions, $\theta$ represents an angle, $\phi$ a dihedral or improper angle, and $K$, $r_0$, $\theta_0$, $K_{ub}$, $r_{ub}$, $n$ and $d$ are parameters that are displayed in tables \ref{intramol} and \ref{intramol2}.

\begin{table}[h!]
\centering
\small
  \caption{Long range interaction parameters}
  \label{long}
  \begin{tabular*}{0.5\textwidth}{@{\extracolsep{\fill}}l}
    \hline
    Coulombic-Lennard Jones \\
\end{tabular*}
  \begin{tabular*}{0.5\textwidth}{@{\extracolsep{\fill}}llll}
     & $q$ (e)& $\epsilon$($10^{-3}$eV) & $\sigma$(\AA)\\
    \hline
Zn        & 0.354   & 0.542 & 1.96  \\
Zn$^*$       & 0.088  &  0  & -   \\
N         &  0 & 7.376 & 3.25   \\
N$^*$        &  -0.42  & 0 &  - \\
C$_1$        &  0.277 & 3.73 & 3.4   \\
H$_1$        &  0.114 & 0.681 &2.47   \\
C$_2$        &  -0.066 & 3.731 & 3.4    \\
H$_2$        &  0.114 & 0.651 & 2.51    \\
C$_{DMF}$        &  0.45 & 4.215 &  3.7   \\
O$_{DMF}$        &  -0.5 & 9.467  & 2.96   \\
N$_{DMF}$        &  -0.57 & 6.744 &  3.2   \\
H$_{DMF}$        &  0.06 & 0.675 & 2.2 \\
CH$_{3DMF}$     & 0.28  & 6.744 &  3.8   \\
    \hline
    Morse & & & \\
    & $D_0$(eV) & $r_0$(\AA) & $\alpha$(\AA$^{-1}$)\\
    \hline
Zn-N     & 0.2  & 2.0 &  4.0   \\
Zn-O$_{(DMF)}$     & 0.5  & 2.1 &  4.0   \\
    \hline
    \vspace{4cm}

 \end{tabular*}
\end{table}

\begin{table}[h!]
\centering
\small
  \caption{Intramolecular interaction parameters (part 1). Species marked with $^*$ represent dummy atoms.}
  \label{intramol}
  \begin{tabular*}{0.65\textwidth}{@{\extracolsep{\fill}}l}
    \hline
    Bonds \\
\end{tabular*}
  \begin{tabular*}{0.65\textwidth}{@{\extracolsep{\fill}}lll}
     & $K$(eV\AA$^{-2}$) & $r_0$(\AA) \\
    \hline
Zn Zn$^*$        & 23.41   & 0.9   \\
N C$_1$        & 14.62   & 1.355   \\
N C$_2$        & 12.55   & 1.386   \\
N N$^*$        & 23.41   & 0.5   \\
C$_1$ H$_1$        & 16.03   & 1.088   \\
C$_2$ H$_2$        & 16.03   & 1.088   \\
C$_2$ C$_2$        & 17.44   & 1.377   \\
Zn$^*$ Zn$^*$        & 23.41   & 1.47   \\
C H$_{DMF}$ & 13.74 & 1.123 \\
C O$_{DMF}$ & 28.19 & 1.23 \\
C N$_{DMF}$ & 18.65  & 1.33 \\
N CH$_{3DMF}$ & 10.41 & 1.44 \\
    \end{tabular*}
    \begin{tabular*}{0.65\textwidth}{@{\extracolsep{\fill}}l}
    \hline
       Angles \\
\end{tabular*}
\begin{tabular*}{0.65\textwidth}{@{\extracolsep{\fill}}lllll}
     & $K$(eV$^{\circ-2}$) & $\theta_0$($^{\circ}$) & $K_{ub}$(eV\AA$^{-2}$) & $r_{ub}$(\AA) \\
    \hline
Zn$^*$ Zn Zn$^*$        & 2.384   & 109.5  & 0 & 0 \\
C$_1$ N C$_2$        & 2.008   & 106.25  & 4.841 & 2.193  \\
C$_1$ N N$^*$        & 0.625   & 126.85  & 0 & 0 \\
C$_2$ N N$^*$      & 0.492   & 126.95  & 0 & 0 \\
N C$_1$ N        & 1.402   & 111.17  & 4.655 & 2.236 \\
N C$_1$ H$_1$        & 1.694   & 124.2  & 0.886 & 2.16 \\
N C$_2$ H$_2$   & 1.369   & 121.32  & 0.886 & 2.16 \\
N C$_2$ C$_2$        & 1.456   & 108  & 4.295 & 2.235 \\
H$_2$ C$_2$ C$_2$        & 130.03  & 130.03  & 0.641 &  2.236 \\
Zn Zn$^*$ Zn$^*$        & 2.384   & 35.5  & 0 & 0  \\
Zn$^*$ Zn$^*$ Zn$^*$        & 2.384   & 60  & 0 & 0 \\
CH$_3$ N CH$_{3DMF}$ & 2.17 & 121 & 0 & 0 \\
CH$_3$ N C$_{DMF}$ & 2.17 & 120 & 0 & 0 \\
N C H$_{DMF}$ & 1.907 & 114.5 & 0 & 0 \\
O C N$_{DMF}$ &  3.25 & 123 & 0 & 0 \\
H C O$_{DMF}$ & 1.907 & 122.5 & 0 & 0 \\
\hline
\vspace{4cm}
\end{tabular*} 
\end{table}

\begin{table}[h!]
\centering
\small
  \caption{Intramolecular interaction parameters (part 2)}
  \label{intramol2}
     \begin{tabular*}{0.6\textwidth}{@{\extracolsep{\fill}}l}
    \hline
    Dihedrals \\
\end{tabular*}
\begin{tabular*}{0.6\textwidth}{@{\extracolsep{\fill}}llll}
     & $K$(eV) & $n$ & $d$($^{\circ}$) \\
    \hline
N C$_1$ N C$_2$        & 0.467   & 2 & 180  \\
N C$_1$ N N$^*$           & 0.0266  & 2 & 180  \\
N C$_2$ C$_2$ N        & 0.665   & 2 & 180 \\
N C$_2$ C$_2$ H$_2$    & 0.154   & 2 & 180  \\
C$_1$ N C$_2$ H$_2$    & 0.158   & 2 & 180  \\
C$_1$ N C$_2$ C$_2$    & 0.288   & 2 & 180  \\
H$_1$ C$_1$ N C$_2$    & 0.158   & 2 & 180  \\
H$_1$ C$_1$ N N$^*$       & 0.01    & 2 & 180  \\
C$_2$ C$_2$ N N$^*$       & 0.061   & 2 & 180  \\
H$_2$ C$_2$ N N$^*$       & 0.0458  & 2 & 180  \\
H$_2$ C$_2$ C$_2$ H$_2$& 0.015   & 2 & 180  \\
H C N CH$_3$$_{DMF}$ & 0.12    & 2 & 180  \\
O C N CH$_3$$_{DMF}$ & 0.12    & 2 & 180  \\
\end{tabular*}
    \begin{tabular*}{0.6\textwidth}{@{\extracolsep{\fill}}l}
    \hline
    Impropers \\
\end{tabular*}
\begin{tabular*}{0.6\textwidth}{@{\extracolsep{\fill}}llll}
     & $K$(eV) & $d$ & $n$  \\
    \hline
N  C$_1$ C$_2$ N$^*$      & 0.152   & -1  & 2\\
C$_1$ N N H$_1$        & 0.152   & -1  & 2\\
    \hline
 \end{tabular*}
\end{table}

\bibliography{magnify}

\providecommand{\latin}[1]{#1}
\makeatletter
\providecommand{\doi}
  {\begingroup\let\do\@makeother\dospecials
  \catcode`\{=1 \catcode`\}=2 \doi@aux}
\providecommand{\doi@aux}[1]{\endgroup\texttt{#1}}
\makeatother
\providecommand*\mcitethebibliography{\thebibliography}
\csname @ifundefined\endcsname{endmcitethebibliography}  {\let\endmcitethebibliography\endthebibliography}{}
\begin{mcitethebibliography}{58}
\providecommand*\natexlab[1]{#1}
\providecommand*\mciteSetBstSublistMode[1]{}
\providecommand*\mciteSetBstMaxWidthForm[2]{}
\providecommand*\mciteBstWouldAddEndPuncttrue
  {\def\EndOfBibitem{\unskip.}}
\providecommand*\mciteBstWouldAddEndPunctfalse
  {\let\EndOfBibitem\relax}
\providecommand*\mciteSetBstMidEndSepPunct[3]{}
\providecommand*\mciteSetBstSublistLabelBeginEnd[3]{}
\providecommand*\EndOfBibitem{}
\mciteSetBstSublistMode{f}
\mciteSetBstMaxWidthForm{subitem}{(\alph{mcitesubitemcount})}
\mciteSetBstSublistLabelBeginEnd
  {\mcitemaxwidthsubitemform\space}
  {\relax}
  {\relax}

\bibitem[Chen \latin{et~al.}(2014)Chen, Yang, Zhu, and Xia]{Chen2014}
Chen,~B.; Yang,~Z.; Zhu,~Y.; Xia,~Y. Zeolitic imidazolate framework materials: recent progress in synthesis and applications. \emph{J. Mater. Chem. A} \textbf{2014}, \emph{2}, 16811–16831\relax
\mciteBstWouldAddEndPuncttrue
\mciteSetBstMidEndSepPunct{\mcitedefaultmidpunct}
{\mcitedefaultendpunct}{\mcitedefaultseppunct}\relax
\EndOfBibitem
\bibitem[Ortiz \latin{et~al.}(2014)Ortiz, Freitas, Boutin, Fuchs, and Coudert]{Ortiz2014}
Ortiz,~A.~U.; Freitas,~A.~P.; Boutin,~A.; Fuchs,~A.~H.; Coudert,~F.-X. What makes zeolitic imidazolate frameworks hydrophobic or hydrophilic? The impact of geometry and functionalization on water adsorption. \emph{Phys. Chem. Chem. Phys.} \textbf{2014}, \emph{16}, 9940–9949\relax
\mciteBstWouldAddEndPuncttrue
\mciteSetBstMidEndSepPunct{\mcitedefaultmidpunct}
{\mcitedefaultendpunct}{\mcitedefaultseppunct}\relax
\EndOfBibitem
\bibitem[Iacomi and Maurin(2021)Iacomi, and Maurin]{Iacomi2021}
Iacomi,~P.; Maurin,~G. ResponZIF Structures: Zeolitic Imidazolate Frameworks as Stimuli-Responsive Materials. \emph{ACS Applied Materials \& Interfaces} \textbf{2021}, \emph{13}, 50602–50642\relax
\mciteBstWouldAddEndPuncttrue
\mciteSetBstMidEndSepPunct{\mcitedefaultmidpunct}
{\mcitedefaultendpunct}{\mcitedefaultseppunct}\relax
\EndOfBibitem
\bibitem[Park \latin{et~al.}(2006)Park, Ni, C{\^{o}}t{\'{e}}, Choi, Huang, Uribe-Romo, Chae, O'Keeffe, and Yaghi]{Park2006}
Park,~K.~S.; Ni,~Z.; C{\^{o}}t{\'{e}},~A.~P.; Choi,~J.~Y.; Huang,~R.; Uribe-Romo,~F.~J.; Chae,~H.~K.; O'Keeffe,~M.; Yaghi,~O.~M. Exceptional chemical and thermal stability of zeolitic imidazolate frameworks. \emph{Proceedings of the National Academy of Sciences} \textbf{2006}, \emph{103}, 10186--10191\relax
\mciteBstWouldAddEndPuncttrue
\mciteSetBstMidEndSepPunct{\mcitedefaultmidpunct}
{\mcitedefaultendpunct}{\mcitedefaultseppunct}\relax
\EndOfBibitem
\bibitem[Yu \latin{et~al.}(2020)Yu, Qiao, Bumstead, Bennett, Yue, and Tao]{Yu2020}
Yu,~Y.; Qiao,~A.; Bumstead,~A.~M.; Bennett,~T.~D.; Yue,~Y.; Tao,~H. Impact of 1-Methylimidazole on Crystal Formation, Phase Transitions, and Glass Formation in a Zeolitic Imidazolate Framework. \emph{Crystal Growth \& Design} \textbf{2020}, \emph{20}, 6528–6534\relax
\mciteBstWouldAddEndPuncttrue
\mciteSetBstMidEndSepPunct{\mcitedefaultmidpunct}
{\mcitedefaultendpunct}{\mcitedefaultseppunct}\relax
\EndOfBibitem
\bibitem[Kaneti \latin{et~al.}(2017)Kaneti, Dutta, Hossain, Shiddiky, Tung, Shieh, Tsung, Wu, and Yamauchi]{Kaneti2017}
Kaneti,~Y.~V.; Dutta,~S.; Hossain,~M. S.~A.; Shiddiky,~M. J.~A.; Tung,~K.; Shieh,~F.; Tsung,~C.; Wu,~K.~C.; Yamauchi,~Y. Strategies for Improving the Functionality of Zeolitic Imidazolate Frameworks: Tailoring Nanoarchitectures for Functional Applications. \emph{Advanced Materials} \textbf{2017}, \emph{29}\relax
\mciteBstWouldAddEndPuncttrue
\mciteSetBstMidEndSepPunct{\mcitedefaultmidpunct}
{\mcitedefaultendpunct}{\mcitedefaultseppunct}\relax
\EndOfBibitem
\bibitem[Lewis \latin{et~al.}(2009)Lewis, Ruiz-Salvador, Gómez, Rodriguez-Albelo, Coudert, Slater, Cheetham, and Mellot-Draznieks]{Lewis2009}
Lewis,~D.~W.; Ruiz-Salvador,~A.~R.; Gómez,~A.; Rodriguez-Albelo,~L.~M.; Coudert,~F.-X.; Slater,~B.; Cheetham,~A.~K.; Mellot-Draznieks,~C. Zeolitic imidazole frameworks: structural and energetics trends compared with their zeolite analogues. \emph{CrystEngComm} \textbf{2009}, \emph{11}, 2272\relax
\mciteBstWouldAddEndPuncttrue
\mciteSetBstMidEndSepPunct{\mcitedefaultmidpunct}
{\mcitedefaultendpunct}{\mcitedefaultseppunct}\relax
\EndOfBibitem
\bibitem[Cheetham \latin{et~al.}(2018)Cheetham, Kieslich, and Yeung]{Cheetham2018}
Cheetham,~A.~K.; Kieslich,~G.; Yeung,~H. H.-M. Thermodynamic and Kinetic Effects in the Crystallization of Metal–Organic Frameworks. \emph{Accounts of Chemical Research} \textbf{2018}, \emph{51}, 659–667\relax
\mciteBstWouldAddEndPuncttrue
\mciteSetBstMidEndSepPunct{\mcitedefaultmidpunct}
{\mcitedefaultendpunct}{\mcitedefaultseppunct}\relax
\EndOfBibitem
\bibitem[Zhang \latin{et~al.}(2019)Zhang, Qiao, Tao, and Yue]{Zhang2019}
Zhang,~J.; Qiao,~A.; Tao,~H.; Yue,~Y. Synthesis, phase transitions and vitrification of the zeolitic imidazolate framework: ZIF-4. \emph{Journal of Non-Crystalline Solids} \textbf{2019}, \emph{525}, 119665\relax
\mciteBstWouldAddEndPuncttrue
\mciteSetBstMidEndSepPunct{\mcitedefaultmidpunct}
{\mcitedefaultendpunct}{\mcitedefaultseppunct}\relax
\EndOfBibitem
\bibitem[Schr\"{o}der \latin{et~al.}(2013)Schr\"{o}der, Baburin, van W\"{u}llen, Wiebcke, and Leoni]{Schrder2013}
Schr\"{o}der,~C.~A.; Baburin,~I.~A.; van W\"{u}llen,~L.; Wiebcke,~M.; Leoni,~S. Subtle polymorphism of zinc imidazolate frameworks: temperature-dependent ground states in the energy landscape revealed by experiment and theory. \emph{CrystEngComm} \textbf{2013}, \emph{15}, 4036–4040\relax
\mciteBstWouldAddEndPuncttrue
\mciteSetBstMidEndSepPunct{\mcitedefaultmidpunct}
{\mcitedefaultendpunct}{\mcitedefaultseppunct}\relax
\EndOfBibitem
\bibitem[Van~Santen(1984)]{VanSanten1984}
Van~Santen,~R.~A. The Ostwald step rule. \emph{The Journal of Physical Chemistry} \textbf{1984}, \emph{88}, 5768–5769\relax
\mciteBstWouldAddEndPuncttrue
\mciteSetBstMidEndSepPunct{\mcitedefaultmidpunct}
{\mcitedefaultendpunct}{\mcitedefaultseppunct}\relax
\EndOfBibitem
\bibitem[Cardew(2023)]{Cardew2023}
Cardew,~P.~T. Ostwald Rule of Stages -- Myth or Reality? \emph{Crystal Growth \& Design} \textbf{2023}, \emph{23}, 3958–3969\relax
\mciteBstWouldAddEndPuncttrue
\mciteSetBstMidEndSepPunct{\mcitedefaultmidpunct}
{\mcitedefaultendpunct}{\mcitedefaultseppunct}\relax
\EndOfBibitem
\bibitem[Widmer \latin{et~al.}(2019)Widmer, Lampronti, Chibani, Wilson, Anzellini, Farsang, Kleppe, Casati, MacLeod, Redfern, Coudert, and Bennett]{Widmer2019}
Widmer,~R.~N.; Lampronti,~G.~I.; Chibani,~S.; Wilson,~C.~W.; Anzellini,~S.; Farsang,~S.; Kleppe,~A.~K.; Casati,~N. P.~M.; MacLeod,~S.~G.; Redfern,~S. A.~T.; Coudert,~F.-X.; Bennett,~T.~D. Rich Polymorphism of a Metal{\textendash}Organic Framework in Pressure{\textendash}Temperature Space. \emph{Journal of the American Chemical Society} \textbf{2019}, \emph{141}, 9330--9337\relax
\mciteBstWouldAddEndPuncttrue
\mciteSetBstMidEndSepPunct{\mcitedefaultmidpunct}
{\mcitedefaultendpunct}{\mcitedefaultseppunct}\relax
\EndOfBibitem
\bibitem[Bouëssel~du Bourg \latin{et~al.}(2014)Bouëssel~du Bourg, Ortiz, Boutin, and Coudert]{BousselduBourg2014}
Bouëssel~du Bourg,~L.; Ortiz,~A.~U.; Boutin,~A.; Coudert,~F.-X. Thermal and mechanical stability of zeolitic imidazolate frameworks polymorphs. \emph{APL Materials} \textbf{2014}, \emph{2}, 124110\relax
\mciteBstWouldAddEndPuncttrue
\mciteSetBstMidEndSepPunct{\mcitedefaultmidpunct}
{\mcitedefaultendpunct}{\mcitedefaultseppunct}\relax
\EndOfBibitem
\bibitem[Fonseca \latin{et~al.}(2021)Fonseca, Gong, Jiao, and Jiang]{Fonseca2021}
Fonseca,~J.; Gong,~T.; Jiao,~L.; Jiang,~H.-L. Metal{\textendash}organic frameworks ({MOFs}) beyond crystallinity: amorphous {MOFs}, {MOF} liquids and {MOF} glasses. \emph{Journal of Materials Chemistry A} \textbf{2021}, \emph{9}, 10562--10611\relax
\mciteBstWouldAddEndPuncttrue
\mciteSetBstMidEndSepPunct{\mcitedefaultmidpunct}
{\mcitedefaultendpunct}{\mcitedefaultseppunct}\relax
\EndOfBibitem
\bibitem[Henke \latin{et~al.}(2018)Henke, Wharmby, Kieslich, Hante, Schneemann, Wu, Daisenberger, and Cheetham]{Henke2018}
Henke,~S.; Wharmby,~M.~T.; Kieslich,~G.; Hante,~I.; Schneemann,~A.; Wu,~Y.; Daisenberger,~D.; Cheetham,~A.~K. Pore closure in zeolitic imidazolate frameworks under mechanical pressure. \emph{Chemical Science} \textbf{2018}, \emph{9}, 1654–1660\relax
\mciteBstWouldAddEndPuncttrue
\mciteSetBstMidEndSepPunct{\mcitedefaultmidpunct}
{\mcitedefaultendpunct}{\mcitedefaultseppunct}\relax
\EndOfBibitem
\bibitem[Méndez and Semino(2024)Méndez, and Semino]{Mendez2024_2}
Méndez,~E.; Semino,~R. Phase diagram of ZIF-4 from computer simulations. \emph{Journal of Materials Chemistry A} \textbf{2024}, \relax
\mciteBstWouldAddEndPunctfalse
\mciteSetBstMidEndSepPunct{\mcitedefaultmidpunct}
{}{\mcitedefaultseppunct}\relax
\EndOfBibitem
\bibitem[Eddaoudi \latin{et~al.}(2015)Eddaoudi, Sava, Eubank, Adil, and Guillerm]{Eddaoudi2015}
Eddaoudi,~M.; Sava,~D.~F.; Eubank,~J.~F.; Adil,~K.; Guillerm,~V. Zeolite-like metal–organic frameworks (ZMOFs): design, synthesis, and properties. \emph{Chemical Society Reviews} \textbf{2015}, \emph{44}, 228–249\relax
\mciteBstWouldAddEndPuncttrue
\mciteSetBstMidEndSepPunct{\mcitedefaultmidpunct}
{\mcitedefaultendpunct}{\mcitedefaultseppunct}\relax
\EndOfBibitem
\bibitem[Van~Vleet \latin{et~al.}(2018)Van~Vleet, Weng, Li, and Schmidt]{VanVleet2018}
Van~Vleet,~M.~J.; Weng,~T.; Li,~X.; Schmidt,~J. In Situ, Time-Resolved, and Mechanistic Studies of Metal–Organic Framework Nucleation and Growth. \emph{Chemical Reviews} \textbf{2018}, \emph{118}, 3681–3721\relax
\mciteBstWouldAddEndPuncttrue
\mciteSetBstMidEndSepPunct{\mcitedefaultmidpunct}
{\mcitedefaultendpunct}{\mcitedefaultseppunct}\relax
\EndOfBibitem
\bibitem[Guillerm and Maspoch(2019)Guillerm, and Maspoch]{Guillerm2019}
Guillerm,~V.; Maspoch,~D. Geometry Mismatch and Reticular Chemistry: Strategies To Assemble Metal–Organic Frameworks with Non-default Topologies. \emph{Journal of the American Chemical Society} \textbf{2019}, \emph{141}, 16517–16538\relax
\mciteBstWouldAddEndPuncttrue
\mciteSetBstMidEndSepPunct{\mcitedefaultmidpunct}
{\mcitedefaultendpunct}{\mcitedefaultseppunct}\relax
\EndOfBibitem
\bibitem[Freund \latin{et~al.}(2021)Freund, Canossa, Cohen, Yan, Deng, Guillerm, Eddaoudi, Madden, Fairen‐Jimenez, Lyu, Macreadie, Ji, Zhang, Wang, Haase, W\"{o}ll, Zaremba, Andreo, Wuttke, and Diercks]{Freund2021}
Freund,~R.; Canossa,~S.; Cohen,~S.~M.; Yan,~W.; Deng,~H.; Guillerm,~V.; Eddaoudi,~M.; Madden,~D.~G.; Fairen‐Jimenez,~D.; Lyu,~H.; Macreadie,~L.~K.; Ji,~Z.; Zhang,~Y.; Wang,~B.; Haase,~F.; W\"{o}ll,~C.; Zaremba,~O.; Andreo,~J.; Wuttke,~S.; Diercks,~C.~S. 25 Years of Reticular Chemistry. \emph{Angewandte Chemie International Edition} \textbf{2021}, \emph{60}, 23946–23974\relax
\mciteBstWouldAddEndPuncttrue
\mciteSetBstMidEndSepPunct{\mcitedefaultmidpunct}
{\mcitedefaultendpunct}{\mcitedefaultseppunct}\relax
\EndOfBibitem
\bibitem[Wang \latin{et~al.}(2022)Wang, Pei, Kalmutzki, Yang, and Yaghi]{Wang2022}
Wang,~H.; Pei,~X.; Kalmutzki,~M.~J.; Yang,~J.; Yaghi,~O.~M. Large Cages of Zeolitic Imidazolate Frameworks. \emph{Accounts of Chemical Research} \textbf{2022}, \emph{55}, 707–721\relax
\mciteBstWouldAddEndPuncttrue
\mciteSetBstMidEndSepPunct{\mcitedefaultmidpunct}
{\mcitedefaultendpunct}{\mcitedefaultseppunct}\relax
\EndOfBibitem
\bibitem[Barsukova \latin{et~al.}(2023)Barsukova, Sapianik, Guillerm, Shkurenko, Shaikh, Parvatkar, Bhatt, Bonneau, Alhaji, Shekhah, Balestra, Semino, Maurin, and Eddaoudi]{Barsukova2023}
Barsukova,~M.; Sapianik,~A.; Guillerm,~V.; Shkurenko,~A.; Shaikh,~A.~C.; Parvatkar,~P.; Bhatt,~P.~M.; Bonneau,~M.; Alhaji,~A.; Shekhah,~O.; Balestra,~S. R.~G.; Semino,~R.; Maurin,~G.; Eddaoudi,~M. Face-directed assembly of tailored isoreticular MOFs using centring structure-directing agents. \emph{Nature Synthesis} \textbf{2023}, \emph{3}, 33–46\relax
\mciteBstWouldAddEndPuncttrue
\mciteSetBstMidEndSepPunct{\mcitedefaultmidpunct}
{\mcitedefaultendpunct}{\mcitedefaultseppunct}\relax
\EndOfBibitem
\bibitem[Castillo-Blas \latin{et~al.}(2024)Castillo-Blas, Chester, Keen, and Bennett]{CastilloBlas2024}
Castillo-Blas,~C.; Chester,~A.~M.; Keen,~D.~A.; Bennett,~T.~D. Thermally activated structural phase transitions and processes in metal–organic frameworks. \emph{Chemical Society Reviews} \textbf{2024}, \emph{53}, 3606–3629\relax
\mciteBstWouldAddEndPuncttrue
\mciteSetBstMidEndSepPunct{\mcitedefaultmidpunct}
{\mcitedefaultendpunct}{\mcitedefaultseppunct}\relax
\EndOfBibitem
\bibitem[Mu \latin{et~al.}(2024)Mu, Wang, Gao, Wu, Shi, and Dong]{Mu2024}
Mu,~K.; Wang,~J.; Gao,~M.; Wu,~Y.; Shi,~Q.; Dong,~J. Template Role of Long Alkyl-Chain Amides in the Synthesis of Zeolitic Imidazolate Frameworks. \emph{ACS Omega} \textbf{2024}, \emph{9}, 34777–34786\relax
\mciteBstWouldAddEndPuncttrue
\mciteSetBstMidEndSepPunct{\mcitedefaultmidpunct}
{\mcitedefaultendpunct}{\mcitedefaultseppunct}\relax
\EndOfBibitem
\bibitem[Lee \latin{et~al.}(2023)Lee, Nam, Yang, and Choe]{Lee2023}
Lee,~S.; Nam,~D.; Yang,~D.~C.; Choe,~W. Unveiling Hidden Zeolitic Imidazolate Frameworks Guided by Intuition‐Based Geometrical Factors. \emph{Small} \textbf{2023}, \emph{19}\relax
\mciteBstWouldAddEndPuncttrue
\mciteSetBstMidEndSepPunct{\mcitedefaultmidpunct}
{\mcitedefaultendpunct}{\mcitedefaultseppunct}\relax
\EndOfBibitem
\bibitem[Méndez and Semino(2024)Méndez, and Semino]{Mendez2024}
Méndez,~E.; Semino,~R. Microscopic mechanism of thermal amorphization of ZIF-4 and melting of ZIF-zni revealed via molecular dynamics and machine learning techniques. \emph{Journal of Materials Chemistry A} \textbf{2024}, \emph{12}, 4572–4582\relax
\mciteBstWouldAddEndPuncttrue
\mciteSetBstMidEndSepPunct{\mcitedefaultmidpunct}
{\mcitedefaultendpunct}{\mcitedefaultseppunct}\relax
\EndOfBibitem
\bibitem[Du \latin{et~al.}(2024)Du, Li, Ganisetti, Bauchy, Yue, and Smedskjaer]{Du2024}
Du,~T.; Li,~S.; Ganisetti,~S.; Bauchy,~M.; Yue,~Y.; Smedskjaer,~M.~M. Deciphering the controlling factors for phase transitions in zeolitic imidazolate frameworks. \emph{National Science Review} \textbf{2024}, \emph{11}, nwae023\relax
\mciteBstWouldAddEndPuncttrue
\mciteSetBstMidEndSepPunct{\mcitedefaultmidpunct}
{\mcitedefaultendpunct}{\mcitedefaultseppunct}\relax
\EndOfBibitem
\bibitem[Shi \latin{et~al.}(2024)Shi, Liu, Yue, Arramel, and Li]{Shi2024}
Shi,~Z.; Liu,~B.; Yue,~Y.; Arramel,~A.; Li,~N. Unraveling medium‐range order and melting mechanism of ZIF‐4 under high temperature. \emph{Journal of the American Ceramic Society} \textbf{2024}, \emph{107}, 3845–3856\relax
\mciteBstWouldAddEndPuncttrue
\mciteSetBstMidEndSepPunct{\mcitedefaultmidpunct}
{\mcitedefaultendpunct}{\mcitedefaultseppunct}\relax
\EndOfBibitem
\bibitem[Castel \latin{et~al.}(2024)Castel, André, Edwards, Evans, and Coudert]{Castel2024}
Castel,~N.; André,~D.; Edwards,~C.; Evans,~J.~D.; Coudert,~F.-X. Machine learning interatomic potentials for amorphous zeolitic imidazolate frameworks. \emph{Digital Discovery} \textbf{2024}, \emph{3}, 355–368\relax
\mciteBstWouldAddEndPuncttrue
\mciteSetBstMidEndSepPunct{\mcitedefaultmidpunct}
{\mcitedefaultendpunct}{\mcitedefaultseppunct}\relax
\EndOfBibitem
\bibitem[Yoneya \latin{et~al.}(2015)Yoneya, Tsuzuki, and Aoyagi]{Yoneya2015}
Yoneya,~M.; Tsuzuki,~S.; Aoyagi,~M. Simulation of metal{\textendash}organic framework self-assembly. \emph{Phys. Chem. Chem. Phys.} \textbf{2015}, \emph{17}, 8649--8652\relax
\mciteBstWouldAddEndPuncttrue
\mciteSetBstMidEndSepPunct{\mcitedefaultmidpunct}
{\mcitedefaultendpunct}{\mcitedefaultseppunct}\relax
\EndOfBibitem
\bibitem[Biswal and Kusalik(2016)Biswal, and Kusalik]{Biswal2016}
Biswal,~D.; Kusalik,~P.~G. Probing Molecular Mechanisms of Self-Assembly in Metal{\textendash}Organic Frameworks. \emph{{ACS} Nano} \textbf{2016}, \emph{11}, 258--268\relax
\mciteBstWouldAddEndPuncttrue
\mciteSetBstMidEndSepPunct{\mcitedefaultmidpunct}
{\mcitedefaultendpunct}{\mcitedefaultseppunct}\relax
\EndOfBibitem
\bibitem[Biswal and Kusalik(2017)Biswal, and Kusalik]{Biswal2017}
Biswal,~D.; Kusalik,~P.~G. Molecular simulations of self-assembly processes in metal-organic frameworks: Model dependence. \emph{J. Chem. Phys.} \textbf{2017}, \emph{147}, 044702\relax
\mciteBstWouldAddEndPuncttrue
\mciteSetBstMidEndSepPunct{\mcitedefaultmidpunct}
{\mcitedefaultendpunct}{\mcitedefaultseppunct}\relax
\EndOfBibitem
\bibitem[Col{\'{o}}n \latin{et~al.}(2019)Col{\'{o}}n, Guo, Antony, Hoffmann, and de~Pablo]{Colon2019}
Col{\'{o}}n,~Y.~J.; Guo,~A.~Z.; Antony,~L.~W.; Hoffmann,~K.~Q.; de~Pablo,~J.~J. Free energy of metal-organic framework self-assembly. \emph{J. Chem. Phys.} \textbf{2019}, \emph{150}, 104502\relax
\mciteBstWouldAddEndPuncttrue
\mciteSetBstMidEndSepPunct{\mcitedefaultmidpunct}
{\mcitedefaultendpunct}{\mcitedefaultseppunct}\relax
\EndOfBibitem
\bibitem[Kollias \latin{et~al.}(2019)Kollias, Cantu, Tubbs, Rousseau, Glezakou, and Salvalaglio]{Kollias2019}
Kollias,~L.; Cantu,~D.~C.; Tubbs,~M.~A.; Rousseau,~R.; Glezakou,~V.-A.; Salvalaglio,~M. Molecular Level Understanding of the Free Energy Landscape in Early Stages of Metal{\textendash}Organic Framework Nucleation. \emph{J. Am. Chem. Soc.} \textbf{2019}, \emph{141}, 6073--6081\relax
\mciteBstWouldAddEndPuncttrue
\mciteSetBstMidEndSepPunct{\mcitedefaultmidpunct}
{\mcitedefaultendpunct}{\mcitedefaultseppunct}\relax
\EndOfBibitem
\bibitem[Wells \latin{et~al.}(2019)Wells, Cessford, Seaton, and D\"{u}ren]{Wells2019}
Wells,~S.~A.; Cessford,~N.~F.; Seaton,~N.~A.; D\"{u}ren,~T. Early stages of phase selection in MOF formation observed in molecular Monte Carlo simulations. \emph{RSC Advances} \textbf{2019}, \emph{9}, 14382–14390\relax
\mciteBstWouldAddEndPuncttrue
\mciteSetBstMidEndSepPunct{\mcitedefaultmidpunct}
{\mcitedefaultendpunct}{\mcitedefaultseppunct}\relax
\EndOfBibitem
\bibitem[Filez \latin{et~al.}(2021)Filez, Caratelli, Rivera-Torrente, Muniz-Miranda, Hoek, Altelaar, Heck, Van~Speybroeck, and Weckhuysen]{Filez2021}
Filez,~M.; Caratelli,~C.; Rivera-Torrente,~M.; Muniz-Miranda,~F.; Hoek,~M.; Altelaar,~M.; Heck,~A.~J.; Van~Speybroeck,~V.; Weckhuysen,~B.~M. Elucidation of the pre-nucleation phase directing metal-organic framework formation. \emph{Cell Reports Physical Science} \textbf{2021}, \emph{2}, 100680\relax
\mciteBstWouldAddEndPuncttrue
\mciteSetBstMidEndSepPunct{\mcitedefaultmidpunct}
{\mcitedefaultendpunct}{\mcitedefaultseppunct}\relax
\EndOfBibitem
\bibitem[Balestra and Semino(2022)Balestra, and Semino]{Balestra2022}
Balestra,~S. R.~G.; Semino,~R. Computer simulation of the early stages of self-assembly and thermal decomposition of {ZIF}-8. \emph{The Journal of Chemical Physics} \textbf{2022}, \emph{157}, 184502\relax
\mciteBstWouldAddEndPuncttrue
\mciteSetBstMidEndSepPunct{\mcitedefaultmidpunct}
{\mcitedefaultendpunct}{\mcitedefaultseppunct}\relax
\EndOfBibitem
\bibitem[Balestra \latin{et~al.}(2023)Balestra, Martínez-Haya, Cruz-Hernández, Lewis, Woodley, Semino, Maurin, Ruiz-Salvador, and Hamad]{Balestra2023}
Balestra,~S. R.~G.; Martínez-Haya,~B.; Cruz-Hernández,~N.; Lewis,~D.~W.; Woodley,~S.~M.; Semino,~R.; Maurin,~G.; Ruiz-Salvador,~A.~R.; Hamad,~S. Nucleation of zeolitic imidazolate frameworks: from molecules to nanoparticles. \emph{Nanoscale} \textbf{2023}, \emph{15}, 3504–3519\relax
\mciteBstWouldAddEndPuncttrue
\mciteSetBstMidEndSepPunct{\mcitedefaultmidpunct}
{\mcitedefaultendpunct}{\mcitedefaultseppunct}\relax
\EndOfBibitem
\bibitem[Su and Ahlquist(2021)Su, and Ahlquist]{Su2021}
Su,~H.; Ahlquist,~M. S.~G. Nonbonded Zr4+ and Hf4+ Models for Simulations of Condensed Phase Metal–Organic Frameworks. \emph{The Journal of Physical Chemistry C} \textbf{2021}, \emph{125}, 6471–6478\relax
\mciteBstWouldAddEndPuncttrue
\mciteSetBstMidEndSepPunct{\mcitedefaultmidpunct}
{\mcitedefaultendpunct}{\mcitedefaultseppunct}\relax
\EndOfBibitem
\bibitem[Jawahery \latin{et~al.}(2019)Jawahery, Rampal, Moosavi, Witman, and Smit]{Jawahery2019}
Jawahery,~S.; Rampal,~N.; Moosavi,~S.~M.; Witman,~M.; Smit,~B. Ab Initio Flexible Force Field for Metal–Organic Frameworks Using Dummy Model Coordination Bonds. \emph{Journal of Chemical Theory and Computation} \textbf{2019}, \emph{15}, 3666–3677\relax
\mciteBstWouldAddEndPuncttrue
\mciteSetBstMidEndSepPunct{\mcitedefaultmidpunct}
{\mcitedefaultendpunct}{\mcitedefaultseppunct}\relax
\EndOfBibitem
\bibitem[Barducci \latin{et~al.}(2008)Barducci, Bussi, and Parrinello]{Barducci2008}
Barducci,~A.; Bussi,~G.; Parrinello,~M. Well-Tempered Metadynamics: A Smoothly Converging and Tunable Free-Energy Method. \emph{Physical Review Letters} \textbf{2008}, \emph{100}, 020603\relax
\mciteBstWouldAddEndPuncttrue
\mciteSetBstMidEndSepPunct{\mcitedefaultmidpunct}
{\mcitedefaultendpunct}{\mcitedefaultseppunct}\relax
\EndOfBibitem
\bibitem[Ishiguro \latin{et~al.}(1990)Ishiguro, Miyauchi, and Ozutumi]{Ishiguro1990}
Ishiguro,~S.-i.; Miyauchi,~M.; Ozutumi,~K. Thermodynamics of formation of binary and ternary complexes of zinc(II) with halide and thiocyanate ions and 2,2'--bipyridine in dimethylformamide. \emph{J. Chem. Soc., Dalton Trans.} \textbf{1990}, 2035–2041\relax
\mciteBstWouldAddEndPuncttrue
\mciteSetBstMidEndSepPunct{\mcitedefaultmidpunct}
{\mcitedefaultendpunct}{\mcitedefaultseppunct}\relax
\EndOfBibitem
\bibitem[Volmer and Weber(1926)Volmer, and Weber]{Volmer1926}
Volmer,~M.; Weber,~A. Keimbildung in ubersattigten Gebilden. \emph{Zeitschrift fur Physikalische Chemie} \textbf{1926}, \emph{119U}, 277–301\relax
\mciteBstWouldAddEndPuncttrue
\mciteSetBstMidEndSepPunct{\mcitedefaultmidpunct}
{\mcitedefaultendpunct}{\mcitedefaultseppunct}\relax
\EndOfBibitem
\bibitem[Aduriz-Arrizabalaga \latin{et~al.}(2023)Aduriz-Arrizabalaga, Mercero, De~Sancho, and Lopez]{AdurizArrizabalaga2023}
Aduriz-Arrizabalaga,~J.; Mercero,~J.~M.; De~Sancho,~D.; Lopez,~X. Rules governing metal coordination in A$\beta$–Zn(<scp>ii</scp>) complex models from quantum mechanical calculations. \emph{Physical Chemistry Chemical Physics} \textbf{2023}, \emph{25}, 27618–27627\relax
\mciteBstWouldAddEndPuncttrue
\mciteSetBstMidEndSepPunct{\mcitedefaultmidpunct}
{\mcitedefaultendpunct}{\mcitedefaultseppunct}\relax
\EndOfBibitem
\bibitem[Pucheta \latin{et~al.}(2016)Pucheta, Prim, Gillet, and Farjon]{Pucheta2016}
Pucheta,~J. E.~H.; Prim,~D.; Gillet,~J.~M.; Farjon,~J. Deciphering the Conformational Choreography of Zinc Coordination Complexes with Standard and Novel Proton NMR Techniques Combined with DFT Methods. \emph{ChemPhysChem} \textbf{2016}, \emph{17}, 1034–1045\relax
\mciteBstWouldAddEndPuncttrue
\mciteSetBstMidEndSepPunct{\mcitedefaultmidpunct}
{\mcitedefaultendpunct}{\mcitedefaultseppunct}\relax
\EndOfBibitem
\bibitem[Sola \latin{et~al.}(1991)Sola, Lledos, Duran, and Bertran]{Sola1991}
Sola,~M.; Lledos,~A.; Duran,~M.; Bertran,~J. Anion binding and pentacoordination in zinc(II) complexes. \emph{Inorganic Chemistry} \textbf{1991}, \emph{30}, 2523–2527\relax
\mciteBstWouldAddEndPuncttrue
\mciteSetBstMidEndSepPunct{\mcitedefaultmidpunct}
{\mcitedefaultendpunct}{\mcitedefaultseppunct}\relax
\EndOfBibitem
\bibitem[Piskunov \latin{et~al.}(2017)Piskunov, Maleeva, Fukin, Cherkasov, and Bogomyakov]{Piskunov2017}
Piskunov,~A.~V.; Maleeva,~A.~V.; Fukin,~G.~K.; Cherkasov,~V.~K.; Bogomyakov,~A.~S. Pentacoordinated bis- o -benzosemiquinonato zinc complexes with different N-ligands: Structure and magnetic properties. \emph{Inorganica Chimica Acta} \textbf{2017}, \emph{455}, 213–220\relax
\mciteBstWouldAddEndPuncttrue
\mciteSetBstMidEndSepPunct{\mcitedefaultmidpunct}
{\mcitedefaultendpunct}{\mcitedefaultseppunct}\relax
\EndOfBibitem
\bibitem[Kirchner and Krebs(1987)Kirchner, and Krebs]{Kirchner1987}
Kirchner,~C.; Krebs,~B. Pentacoordinate zinc complexes of imidazole nitrogen donors as structural models for the active site in enzymes: preparation and crystal structures of (.mu.-2, 2’-biimidazole)tetrakis(2, 2’-biimidazole)dizinc(II) tetraperchlorate trihydrate and bis(2, 2’-biimidazole)(formato)zinc(II) perchlorate. \emph{Inorganic Chemistry} \textbf{1987}, \emph{26}, 3569–3576\relax
\mciteBstWouldAddEndPuncttrue
\mciteSetBstMidEndSepPunct{\mcitedefaultmidpunct}
{\mcitedefaultendpunct}{\mcitedefaultseppunct}\relax
\EndOfBibitem
\bibitem[Thompson \latin{et~al.}(2022)Thompson, Aktulga, Berger, Bolintineanu, Brown, Crozier, in~{\textquotesingle}t~Veld, Kohlmeyer, Moore, Nguyen, Shan, Stevens, Tranchida, Trott, and Plimpton]{lammps}
Thompson,~A.~P.; Aktulga,~H.~M.; Berger,~R.; Bolintineanu,~D.~S.; Brown,~W.~M.; Crozier,~P.~S.; in~{\textquotesingle}t~Veld,~P.~J.; Kohlmeyer,~A.; Moore,~S.~G.; Nguyen,~T.~D.; Shan,~R.; Stevens,~M.~J.; Tranchida,~J.; Trott,~C.; Plimpton,~S.~J. {LAMMPS} - a flexible simulation tool for particle-based materials modeling at the atomic, meso, and continuum scales. \emph{Computer Physics Communications} \textbf{2022}, \emph{271}, 108171\relax
\mciteBstWouldAddEndPuncttrue
\mciteSetBstMidEndSepPunct{\mcitedefaultmidpunct}
{\mcitedefaultendpunct}{\mcitedefaultseppunct}\relax
\EndOfBibitem
\bibitem[Tribello \latin{et~al.}(2014)Tribello, Bonomi, Branduardi, Camilloni, and Bussi]{Tribello2014}
Tribello,~G.~A.; Bonomi,~M.; Branduardi,~D.; Camilloni,~C.; Bussi,~G. PLUMED 2: New feathers for an old bird. \emph{Computer Physics Communications} \textbf{2014}, \emph{185}, 604–613\relax
\mciteBstWouldAddEndPuncttrue
\mciteSetBstMidEndSepPunct{\mcitedefaultmidpunct}
{\mcitedefaultendpunct}{\mcitedefaultseppunct}\relax
\EndOfBibitem
\bibitem[Chalaris and Samios(2000)Chalaris, and Samios]{Chalaris2000}
Chalaris,~M.; Samios,~J. Systematic molecular dynamics studies of liquid N, N-dimethylformamide using optimized rigid force fields: Investigation of the thermodynamic, structural, transport and dynamic properties. \emph{The Journal of Chemical Physics} \textbf{2000}, \emph{112}, 8581–8594\relax
\mciteBstWouldAddEndPuncttrue
\mciteSetBstMidEndSepPunct{\mcitedefaultmidpunct}
{\mcitedefaultendpunct}{\mcitedefaultseppunct}\relax
\EndOfBibitem
\bibitem[Vanommeslaeghe \latin{et~al.}(2009)Vanommeslaeghe, Hatcher, Acharya, Kundu, Zhong, Shim, Darian, Guvench, Lopes, Vorobyov, and Mackerell]{Vanommeslaeghe2009}
Vanommeslaeghe,~K.; Hatcher,~E.; Acharya,~C.; Kundu,~S.; Zhong,~S.; Shim,~J.; Darian,~E.; Guvench,~O.; Lopes,~P.; Vorobyov,~I.; Mackerell,~A.~D. CHARMM general force field: A force field for drug‐like molecules compatible with the CHARMM all‐atom additive biological force fields. \emph{Journal of Computational Chemistry} \textbf{2009}, \emph{31}, 671–690\relax
\mciteBstWouldAddEndPuncttrue
\mciteSetBstMidEndSepPunct{\mcitedefaultmidpunct}
{\mcitedefaultendpunct}{\mcitedefaultseppunct}\relax
\EndOfBibitem
\bibitem[Raiteri \latin{et~al.}(2005)Raiteri, Laio, Gervasio, Micheletti, and Parrinello]{Raiteri2005}
Raiteri,~P.; Laio,~A.; Gervasio,~F.~L.; Micheletti,~C.; Parrinello,~M. Efficient Reconstruction of Complex Free Energy Landscapes by Multiple Walkers Metadynamics. \emph{The Journal of Physical Chemistry B} \textbf{2005}, \emph{110}, 3533–3539\relax
\mciteBstWouldAddEndPuncttrue
\mciteSetBstMidEndSepPunct{\mcitedefaultmidpunct}
{\mcitedefaultendpunct}{\mcitedefaultseppunct}\relax
\EndOfBibitem
\bibitem[Tiwary and Parrinello(2014)Tiwary, and Parrinello]{Tiwary2014}
Tiwary,~P.; Parrinello,~M. A Time-Independent Free Energy Estimator for Metadynamics. \emph{The Journal of Physical Chemistry B} \textbf{2014}, \emph{119}, 736–742\relax
\mciteBstWouldAddEndPuncttrue
\mciteSetBstMidEndSepPunct{\mcitedefaultmidpunct}
{\mcitedefaultendpunct}{\mcitedefaultseppunct}\relax
\EndOfBibitem
\bibitem[Bussi and Tribello(2019)Bussi, and Tribello]{Bussi2019}
Bussi,~G.; Tribello,~G.~A. \emph{Biomolecular Simulations}; Springer New York, 2019; p 529–578\relax
\mciteBstWouldAddEndPuncttrue
\mciteSetBstMidEndSepPunct{\mcitedefaultmidpunct}
{\mcitedefaultendpunct}{\mcitedefaultseppunct}\relax
\EndOfBibitem
\bibitem[Bennett \latin{et~al.}(2011)Bennett, Simoncic, Moggach, Gozzo, Macchi, Keen, Tan, and Cheetham]{Bennett2011_3}
Bennett,~T.~D.; Simoncic,~P.; Moggach,~S.~A.; Gozzo,~F.; Macchi,~P.; Keen,~D.~A.; Tan,~J.-C.; Cheetham,~A.~K. Reversible pressure-induced amorphization of a zeolitic imidazolate framework ({ZIF}-4). \emph{Chemical Communications} \textbf{2011}, \emph{47}, 7983\relax
\mciteBstWouldAddEndPuncttrue
\mciteSetBstMidEndSepPunct{\mcitedefaultmidpunct}
{\mcitedefaultendpunct}{\mcitedefaultseppunct}\relax
\EndOfBibitem
\end{mcitethebibliography}

\end{document}